%% file: main.tex
\documentclass[lettersize,journal]{IEEEtran}
\usepackage{amsmath,amssymb,amsfonts}
\usepackage{algorithmic}
\usepackage{algorithm}
\usepackage{array}
\usepackage[caption=false]{subfig}
\usepackage{textcomp}
\usepackage{stfloats}
\usepackage{url}
\usepackage{verbatim}
\usepackage{graphicx}
\usepackage{cite}

\usepackage{tabularx}
\usepackage{booktabs}
\usepackage[table]{xcolor}
\usepackage{ragged2e}
\usepackage{makecell}
\usepackage{multirow}

\usepackage{CJKutf8}

\usepackage[normalem]{ulem} 

\usepackage{hyperref}
\usepackage{cleveref}

\graphicspath{{images/}{images/section3-design/}{images/section4-setup/}{images/section5-evaluation/}{images/photo/}}

\newif\ifEnableMark
\newif\ifEnableMapping

\EnableMarktrue    
\EnableMappingtrue 

\newcommand{\GlossaryCalibration}[2]{%
    \ifEnableMapping
        #2%
    \else
        \ifEnableMark
            \textcolor{gray}{\sout{#1}}\textcolor{orange}{\textbf{#2}}%
        \else
            #1%
        \fi
    \fi
}

\newcommand{\myname}{CerebroSim}

\newcommand{\smerng}{FlySyn} 

\ifdefined\EnableTodo
    \newcommand{\MyComment}[2]{\textcolor{red}{#1}\textbf{{\begin{CJK*}{UTF8}{gbsn}\textcolor{blue}{ (Comment: #2)}\end{CJK*}}}}
\else
    \newcommand{\MyComment}[2]{#1}
\fi

\def\BibTeX{{\rm B\kern-.05em{\sc i\kern-.025em b}\kern-.08em
    T\kern-.1667em\lower.7ex\hbox{E}\kern-.125emX}}

\usepackage{etoolbox}
\makeatletter
\def\@IEEEBIOskipN{0\baselineskip}
\patchcmd{\IEEEbiography}%
  {\vskip \@IEEEBIOskipN plus 1fil minus 0\baselineskip}%
  {\vskip \@IEEEBIOskipN}%
  {}{}
\makeatother

\begin{document}

\title{\myname{}: Scalable Whole-Brain Simulator \\ at 100-Trillion-Synapse Scale \\ on the LineShine Supercomputer
}

\author{Guangnan Feng,
        Tianxiang Lyu,
        Hao Huang,
        Honghui Liang,
        Jingjing Li,
        Zhiguang Chen,
        and Yutong Lu$^{*}$
\thanks{Manuscript received XXX, 2026; revised XXX, 2026.}
\thanks{Guangnan Feng, Tianxiang Lyu, Hao Huang, Honghui Liang, Jingjing Li, and Zhiguang Chen are with Sun Yat-sen University, Guangzhou 510006, China (e-mail: fenggn7@mail.sysu.edu.cn; lvtx@mail2.sysu.edu.cn; huangh558@mail2.sysu.edu.cn; lianghh33@mail2.sysu.edu.cn; ljj399@mail2.sysu.edu.cn; chenzhg29@mail.sysu.edu.cn).}
\thanks{Yutong Lu is with Sun Yat-sen University, Guangzhou 510006, China, and also with the National Supercomputing Center in Shenzhen, Shenzhen 518055, China, and is the corresponding author (e-mail: luyutong@mail.sysu.edu.cn).}}

\markboth{ }%
{Feng \MakeLowercase{\textit{et al.}}: \myname{}: Scalable Whole-Brain Simulator at 100-Trillion-Synapse Scale}


\maketitle

\input{sections/section0_abstract}

\input{sections/section1_introduction}

\input{sections/section2_background}
\input{sections/section2back_lineshine}
\input{sections/section3_design}
\input{sections/section4_setup}
\input{sections/section5_evaluation}

\input{sections/section6_conclusion}

\input{sections/section93_references}

\input{sections/section95_biography}

\end{document}

%% file: sections/section0_abstract.tex
\begin{abstract}

Building executable brain models is essential for moving neuroscience from description to mechanism and prediction.
Human-brain-scale spiking simulation is constrained by highly irregular communication, multithreaded spike delivery, and the memory cost of sparse connectivity.
We present \myname{}, a scalable framework for whole-brain simulation.
\myname{} combines Delay-aware Spike Broadcast (DSB) for aggregated delay-aware communication, Race-free Synaptic Dynamics Computation (RSDC) for lock/atomic-free multithreaded delivery with HBM-aware optimization, and Sparse Synapse Storage Compression (3SC) for compact indexing with deterministic synapse regeneration.
Using a model derived from magnetic resonance imaging and diffusion-weighted imaging, \myname{} simulates 86 billion neurons and 100 trillion synapses on 18{,}432 nodes across 11.2 million cores of the LineShine Supercomputer, sustaining 24.44~PFlop/s, 91\% weak-scaling efficiency, and 94\% strong-scaling efficiency.
This capability makes biologically constrained human-brain models practical for mechanistic studies of brain disorders and controlled in silico testing of intervention hypotheses.

\end{abstract}


%% file: sections/section1_introduction.tex
\section{Introduction}

\IEEEPARstart{U}{nderstanding} and simulating the human brain have long been regarded as among the most ambitious scientific challenges \cite{yamazaki_human_scale_2021, kandel_2013}.
We are currently witnessing a paradigm shift in brain modeling, moving from static, coarse-grained descriptions---such as mean-field models \cite{pinotsis_neural_2014, byrne_mean-field_2022} and region-level connectivity analyses \cite{avvaru_region-level_2021}---toward the more ambitious goal of large-scale, mechanistic brain simulation at the level of neurons and synapses \cite{fan_brief_2019, markramReconstructionSimulationNeocortical2015}.
While traditional approaches, including neural mass models \cite{deschle_validity_2021} and statistical reconstructions from functional Magnetic Resonance Imaging (fMRI) \cite{glover_overview_2012}, have provided valuable insights into global brain dynamics, they fundamentally remain descriptive models rather than executable simulators.
They capture correlations in neural activity but often lack grounded representations of anatomical details and the underlying mechanisms, such as spike-based communication, synaptic plasticity, and emergent computation.
Bridging this gap requires moving from passive observation of brain signals to the construction of executable brain models that can reproduce, predict, and interact with neural processes.

However, the transition from descriptive models to whole-brain simulation faces significant challenges.
First, high-resolution, multi-scale data remain scarce.
Unlike imaging modalities such as fMRI \cite{glover_overview_2012} or electroencephalography~\cite{mushtaq_electroencephalography_2025}, which provide indirect and often low-temporal-resolution observations, large-scale recordings \cite{paulk_large-scale_2022} of spiking activity with precise anatomical and synaptic detail do not yet scale.
Second, irregular communication patterns and sparse memory accesses introduce substantial inefficiencies in large-scale brain simulations.
Unlike dense numerical workloads, spiking neural activity is inherently event-driven and highly sparse, leading to unpredictable, fine-grained communication across distributed compute nodes.
This irregularity results in poor network utilization, increased latency, and synchronization overhead, particularly when spikes must be routed across distant regions of a simulated brain network.
Moreover, the massive memory residency required for synaptic data serves as a critical bottleneck, while sparse and noncontiguous memory accesses further degrade cache locality and bandwidth efficiency, severely limiting performance on modern hardware architectures.
As a result, existing simulation frameworks often struggle to fully exploit parallelism, which calls for novel data structures, communication strategies, and hardware-aware optimizations to efficiently handle the sparse and asynchronous nature of brain activity.

In this work, we present a scalable framework for whole-brain simulation that addresses these challenges and enables the next generation of computational neuroscience.
\myname{} is not merely a numerical simulator; it is a unified system that integrates data, models, and high-performance computing into a coherent platform for studying the brain.
It is designed to capture fine-grained neural dynamics while scaling to billions of neurons.
Through communication-optimized parallelism and memory-efficient data structures, the framework achieves strong scalability on modern supercomputers.
This capability enables long-horizon simulations with massive neuronal populations, realistic synaptic delays, and realistic brain architecture.

To achieve extreme scalability under sparse, irregular, and delay-constrained whole-brain workloads, we realize three innovations that directly target the dominant communication and memory bottlenecks:

\begin{itemize}
    \item \textbf{Delay-aware Spike Broadcast (DSB).} DSB restructures globally irregular spike communication into aggregated hop-by-hop forwarding over virtual topologies.
    It further converts biological delay windows into schedulable slack for delay-aware transmission and communication-computation overlap, while a lightweight transfer semantic reduces the protocol overhead.

    \item \textbf{Race-free Synaptic Dynamics Computation (RSDC).} RSDC eliminates locks and atomic operations in multithreaded spike delivery through data partitioning, deterministic scheduling, and conflict-free updates.
    We further combine this design with HBM prefetching to mitigate irregular memory accesses and improve memory-bound synaptic processing throughput.

    \item \textbf{Sparse Synapse Storage Compression (3SC).} To overcome the prohibitive memory demands of whole-brain simulation, 3SC introduces a Synapse-Aware Compressed Index (SACI) for compressing the index entries. 
    Furthermore, we propose \smerng{}, an on-the-fly synapse regeneration scheme to fundamentally replace synaptic storage with real-time computation.
\end{itemize}

\input{sections/section3_0fig_overview}

%% file: sections/section3_0fig_overview.tex
\begin{figure*}[ht]
  \centering
  \includegraphics[width=1.0\linewidth]{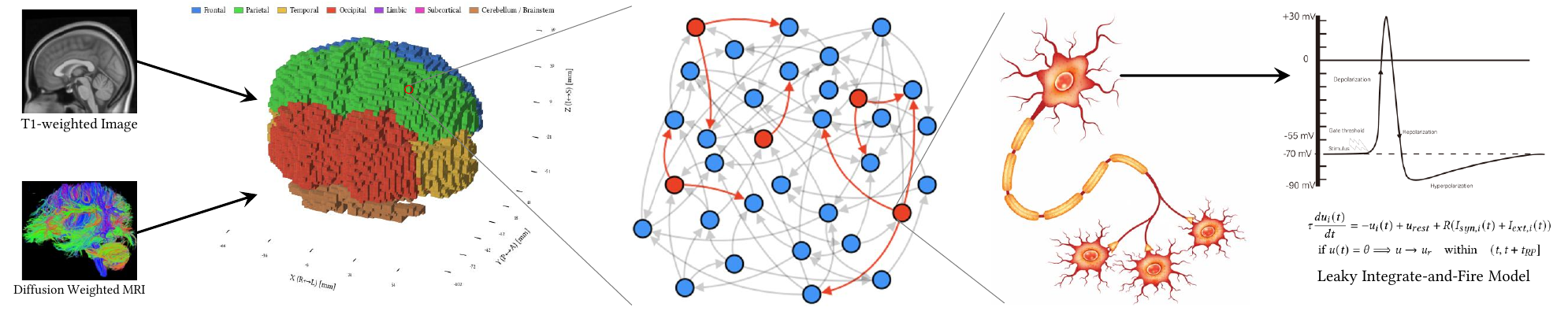}
  \caption{\textbf{\myname{} Simulation Overview.}}
  \label{fig:overview}
\end{figure*}

%% file: sections/section2_background.tex
\section{Background and Related Work on Brain Simulation}

\subsection{Current Scale and Performance}

Recent progress in brain simulation has pushed the field from local circuits toward whole-cortex and human-brain-scale models.
\cite{kuriyama_microscopic_level_2025} demonstrated a mouse whole-cortex simulation with 9 million biophysical neurons and 26 billion synapses on Fugaku, using 145,728 compute nodes and reporting an estimated 7.31 PFlop/s performance, together with strong scaling to more than 150,000 nodes. \cite{lu_simulation_2024} reported a human-brain-scale Digital Brain model with 86 billion neurons and \MyComment{47.8 trillion synapses}{HH：和前面160 trillion synapses的数据有一点差异，看有没有影响？} on 3,503 nodes and 14,012 GPUs.

The computational opportunity behind these advances is the massive parallelism exposed by neuron and synapse updates. \cite{igarashi_future_2025} emphasized that spiking neural network simulation offers abundant parallel work and can benefit from partitioning strategies aligned with cortical and cerebellar structures\cite{lawrenson_mystery_2018}.
\cite{kuriyama_microscopic_level_2025} combined MPI, OpenMP, and vector-level parallelism in a hierarchical mapping from compute nodes to the model structure. \MyComment{\cite{lu_simulation_2024} adopted partitioning and two-level routing to reduce inter-GPU traffic}{HH：和后面有点重复看起来（？）}. Together, these results define the current frontier as the ability to translate biological scale into sustained parallel efficiency on modern supercomputers.

\subsection{Spike Propagation Communication}

Spike propagation in distributed brain simulation is fundamentally a communication problem dominated by massive irregular broadcasts of tiny spike messages with heterogeneous delays.
In distributed execution, inter-process messages carry spike events emitted by pre-synaptic neurons to remote processes that host their post-synaptic targets.
This pattern arises from sparse, heterogeneous, and topologically irregular biological connectivity \cite{fernandez-musolesCommunicationSparsityDistributed2019}, while the different physical lengths of synaptic pathways require the simulator to preserve a delay for each connection rather than delivering all spikes in the same communication step.
Moreover, the extremely low-power operating regime of the brain is reflected in sparse firing activity, so most neurons remain silent in most simulation steps.
\Cref{tab:brain_comm_stats} shows the statistical characteristics of broadcast operations in a whole-brain simulation.

\input{sections/section2_0tab_comm}

To execute this communication pattern on current machines, prior large-scale simulators have developed practical and innovative implementations based on mature message-passing abstractions and machine-aware routing.
\cite{kuriyama_microscopic_level_2025, kunkel_spiking_2014, hinesComparisonNeuronalSpike2011} mapped spike propagation onto \texttt{MPI\_Allgather}, \texttt{MPI\_Allgatherv}, showing that carefully engineered collective-communication workflows can support whole-cortex-scale execution on Fugaku.
In NEST \cite{jordanExtremelyScalableSpiking2018}, instead of every rank receiving spikes from all neurons globally, spikes are packed into buffers and routed between MPI ranks using \texttt{MPI\_Alltoall}, matching the graph nature of neural connectivity.
\cite{lu_simulation_2024} introduced a two-level routing scheme for human-brain-scale deployment, demonstrating that hierarchical traffic organization can reduce inter-GPU pressure while preserving integration with full simulation workflows.
\cite{igarashi_future_2025} further highlighted the importance of partitioning and runtime organization for sustaining efficiency in large spiking simulations.

Brain-simulation communication nevertheless remains statistically misaligned with the dominant hardware and software assumptions of modern supercomputers.
At the hardware level, network links typically approach peak bandwidth only for packets of several kilobytes, often around 4~KB to 8~KB, whereas the 357~B average payload in \Cref{tab:brain_comm_stats} would utilize only about 4.4\% of the link bandwidth under an 8~KB efficiency target.
At the software level, mainstream communication libraries deliver their best scalability through highly optimized structured collectives, such as Alltoall, Allreduce, Allgather, and Bcast, rather than through interfaces that directly express massive irregular fan-out with destination-specific delays\cite{mpich,dragonfly}.
As a result, current implementations must translate biologically required communication semantics into available primitives, which can introduce additional synchronization, traffic reorganization, and limited overlap opportunities at scale.
This mismatch motivates the need for a new communication method that directly targets tiny payloads, massive irregular broadcasts, and delay-aware propagation in distributed brain simulation.

Moreover, the communication volume is not only sparse but also highly skewed, with certain regions of the brain exhibiting significantly higher firing rates and connectivity density.
This leads to load imbalance at the communication level, where a subset of ranks becomes communication hotspots.
Existing MPI collectives are not designed to handle such skew efficiently, and hardware-level optimizations (e.g., topology-aware routing) are often underutilized due to the dynamic nature of spike traffic.

\subsection{Race Conditions in Spike Delivery}

Within a process, spike delivery is difficult to parallelize across threads because concurrent writes to shared state introduce race conditions. The spike traffic is distributed across threads, yet every thread writes to shared post-synaptic states, and a post-synaptic neuron receives synapses from many pre-synaptic neurons. Spikes emitted by different neurons in the same simulation step can therefore reach the same target, and the threads that process them update the same state concurrently, which gives rise to a race condition. Because the colliding writes are determined by the connectivity of the network rather than by the layout of the data, they cannot be avoided by how the incoming spikes are distributed across threads.

These collisions are occasional, and they are scattered across memory rather than clustered. Connectivity in the brain is sparse, with the 86 billion neurons of the human brain connected by 100 trillion synapses, so each neuron connects to only a few thousand others, and only a small fraction of the synapses is active in any simulation step. Although the synapses that share a pre-synaptic neuron are stored contiguously, their post-synaptic targets are spread across the target arrays according to connectivity, so colliding writes are rarely close together in memory. Both the frequency and the location of a collision are therefore determined by connectivity and firing activity rather than by anything a simulator can control in advance.

Existing implementations have explored several practical solutions.
A direct approach is to use mutexes or atomic operations~\cite{lu_simulation_2024} to preserve correctness during concurrent updates.
Other designs use thread-private buffers \cite{kandel_2013} followed by reduction, or adopt lock-free data structures to reduce explicit contention.
These strategies enable shared-memory parallelism in spike delivery, and each offers a different trade-off among implementation simplicity, synchronization cost, and memory overhead.

\subsection{Connectivity Storage and Compression}

Memory footprint is another defining constraint in large-scale brain simulation because biologically realistic models must manage enormous connectivity data.
At the scale of billions of neurons and trillions of synapses, even the modest spike-synapse index translates into a very large memory footprint and heavy memory traffic.
This makes data representation and storage reduction central design choices rather than secondary implementation details \cite{kumbharCoreNEURONOptimizedCompute2019}.

Existing simulators commonly use sparse connectivity representations, such as adjacency lists, compressed-sparse-row- or compressed-sparse-column-like layouts \cite{morrison2005advancing, jordan2020efficient, pronold2022routing}, so that only existing synapses are stored and indexed.
Sparse connection-infrastructure tables have also been explored to reduce metadata for neurons without local targets \cite{kunkelMeetingMemoryChallenges2012}.
These formats improve memory efficiency relative to dense representations and make connectivity traversal straightforward during spike delivery. They also align naturally with partitioned simulation, where each process stores and updates only its local subset of neurons and synapses. Such designs have been instrumental in making very large spiking models executable on current systems. Nevertheless, most existing simulators store synaptic connectivity, weights, delays, and state variables in relatively straightforward data structures without aggressive compression. While this simplifies implementation and enables fast indexing, it fails to exploit the sparsity of biological connectivity, and it limits the maximum model size that can fit in system memory, directly constraining simulation scale. Alternatively, procedural connectivity \cite{knightLargerGPUacceleratedBrain2021} enables on-the-fly synaptic regeneration on GPUs by leveraging the deterministic nature of pseudo-random processes, effectively eliminating explicit connectivity-storage overhead.

Overall, despite substantial progress in simulation scale and parallel performance, current whole-brain simulators still face three tightly coupled challenges: inefficient irregular spike communication, synchronization and locality bottlenecks in multithreaded spike delivery, and prohibitive memory overhead in large-scale connectivity storage. These limitations continue to constrain both scalability and biological fidelity, and therefore motivate the coordinated innovations introduced next in \myname{}.

%% file: sections/section2_0tab_comm.tex
\begin{table}[htb]
\caption{Statistical Characteristics of Broadcast Operations in a Whole-Brain Simulation on 18{,}432 Nodes $\times$ 16 Procs}
\label{tab:brain_comm_stats}
\centering
\small
\begin{tabular}{lrrrr}
\toprule
Metric & Avg. & Max. & Min. & Std. \\
\midrule
\makecell[l]{Bcast msg size (B)} & 357 & 53{,}320 & 0 & 186.4 \\
\makecell[l]{Bcast peers/proc} & 18{,}983.9 & 91{,}856 & 224 & 14{,}249.7 \\
\makecell[l]{Bcast delay (steps)} & 75.7 & 354 & 1 & 38.8 \\
\bottomrule
\end{tabular}
\end{table}

%% file: sections/section2back_lineshine.tex
\section{LineShine Supercomputer}

CerebroSim is developed and evaluated on LineShine \cite{luLineShineOnlineAcceleration2026}, an exascale supercomputer deployed at the National Supercomputing Center in Shenzhen (NSCC-SZ), China.
LineShine comprises 22,680 compute nodes and more than 13 million processor cores.
As shown in \Cref{fig:node_arch}, each compute node integrates two chiplet-based LX2 processors, providing 608 cores and approximately 120.6 TFlop/s of peak FP64 performance per node.

Each LX2 processor contains 304 ARM cores organized into eight NUMA domains, resulting in 16 NUMA domains per node.
Each core incorporates hardware support for both the Scalable Vector Extension (SVE) \cite{stephensARMScalableVector2017} and the Scalable Matrix Extension (SME) \cite{ArmSME}, providing per-core vector and matrix execution capabilities.
This many-core architecture, together with the integrated vector and matrix computing capabilities, provides substantial opportunities for exploiting fine-grained parallelism and diverse computation patterns.

The LX2 adopts a hierarchical HBM–DDR memory architecture that combines high bandwidth with large memory capacity \cite{junHBMHighBandwidth2017}.
Each processor provides 32 GB of on-package HBM with approximately 4 TB/s bandwidth and 256 GB of DDR5 memory, giving each compute node 64 GB of HBM and 512 GB of DDR5.
The HBM can be explicitly managed, enabling bandwidth-sensitive data to be placed in the high-bandwidth tier while larger-capacity data remain in DDR5.
Such a hierarchy provides opportunities to optimize workloads with both high memory-bandwidth demands and large memory footprints.

Compute nodes are interconnected by the LingQi high-speed interconnect.
Each node integrates eight 200 Gbps network interfaces, providing an aggregate network bandwidth of 1.6 Tb/s per node.
The NICs are integrated into the LX2 package and can directly access both HBM and DDR for RDMA communication.
At system scale, LingQi adopts a 2-plane $\times$ 4-rail fat-tree topology, with a measured single-hop latency of 1.07 us and bidirectional inter-node bandwidth of up to 381 GB/s.
These characteristics provide a high-performance communication substrate for large-scale workloads with intensive inter-node data exchange.

\input{sections/section4_0fig_node_arch}


%% file: sections/section4_0fig_node_arch.tex
\begin{figure}[t]
  \centering
  \includegraphics[width=1.0\linewidth]{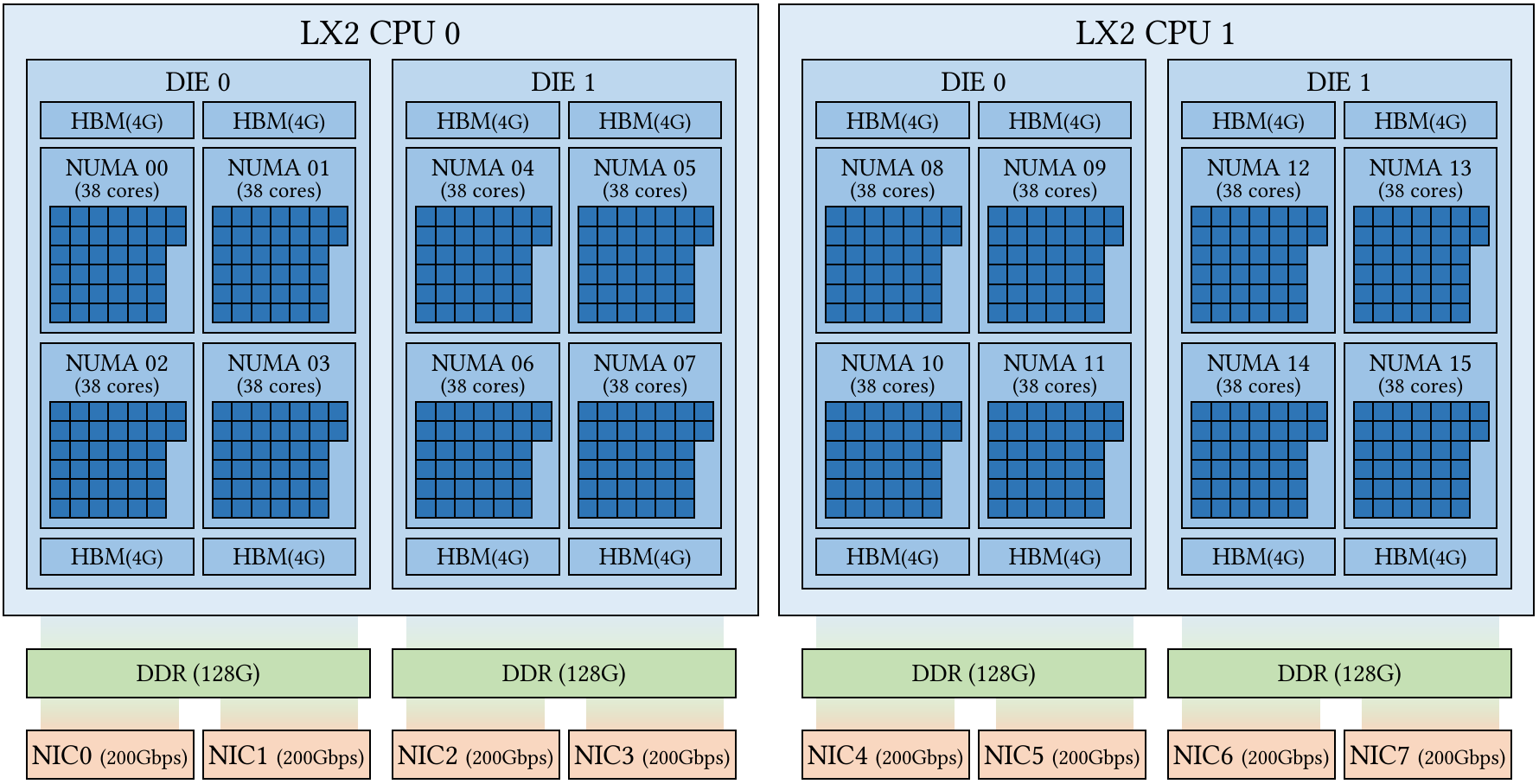}
  \caption{\textbf{Compute Node Architecture of the LineShine Supercomputer.}}
  \label{fig:node_arch} 
\end{figure}

%% file: sections/section3_design.tex
\section{\myname{}: Design and Innovations}

\myname{} enables multi-scale, biologically constrained brain modeling by integrating MRI-derived data with neuron-level dynamics in a unified framework.
As shown in \Cref{fig:overview}, the framework transforms structural information derived from magnetic resonance imaging into a biologically grounded spiking neural network, where voxel-level anatomical and connectivity constraints are systematically translated into neuron- and synapse-level representations.
Building on this framework, we propose the following three innovations.

\subsection{Delay-Aware Spike Broadcast (DSB)}

\input{sections/section3_1_dsb}

\subsection{Race-Free Synaptic Dynamics Computation (RSDC)}
\label{sec:rsdc}

\input{sections/section3_2_rsdc}

\subsection{Sparse Synapse Storage Compression (3SC)}
\label{sec:3sc}

\input{sections/section3_3_saci}

\subsection{On-the-Fly Synapse Regeneration with \smerng{}}
\label{sec:hashmap}

\input{sections/section3_4_flysyn}

%% file: sections/section3_1_dsb.tex
Delay-aware Spike Broadcast targets delay-constrained, tiny-payload, globally irregular spike broadcasts by jointly optimizing forwarding, scheduling, and transport semantics.
Instead of treating each spike as an independent communication event, DSB constructs a reusable forwarding structure that exposes opportunities for aggregation and delay-aware scheduling.

DSB maps massive broadcasts with heterogeneous destinations and delays onto a virtual topology and forwards them hop by hop along static routes, as shown in \Cref{fig:dsb}.
The current implementation supports both configurable multidimensional HyperX \cite{ahnHyperXTopologyRouting2009} and Dragonfly \cite{kimTechnologyDrivenHighlyScalableDragonfly2008, kimCostEfficientDragonflyTopology2009} virtual topologies.
Messages headed to the same next hop are adaptively aggregated into larger transfer units, and relay nodes parse these blocks to determine local delivery and subsequent forwarding before regrouping the surviving entries by next hop.
This next-hop aggregation sharply reduces the number of tiny packets injected into the network and improves effective bandwidth utilization for sparse spike traffic.

\input{sections/section3_1fig_dsb}

DSB does not treat biological delays as passive waiting time, but instead converts them into schedulable slack through step-indexed deferred-send queues.
In the urgent phase, the runtime sends all spikes that must leave in the current communication step at full speed so that communication deadlines are never violated.
In the speculative phase, the runtime advances spikes that belong to future simulation steps.
This speculative transmission means that network movement is completed early while the step at which a message becomes visible to the application remains unchanged.
To prevent excessive speculation from creating queueing pressure for later urgent traffic, DSB throttles further speculative injection according to the number of submitted but unfinished packets.
In both phases, DSB packs as many spikes as possible into packets up to a fixed maximum size, thereby improving bandwidth utilization.
As a result, DSB preserves exact propagation semantics while converting delay windows into opportunities for communication-computation overlap.

For transport semantics, we construct a Stream Data Transfer (SDT) layer.
SDT relies on Verbs send/recv semantics \cite{RDMAawareNetworksProgramming2026}, so the sender does not need the receiver's remote memory address.
Once the receiver has pre-posted receive work requests, the sender can directly advance streaming data transfers.
Compared with two-sided MPI point-to-point communication, this design avoids communicator- and tag-based message-matching overheads.
Compared with rendezvous protocols, it removes the extra handshake before data transfer, and compared with eager protocols, it avoids additional copies associated with unexpected-message buffering \cite{unr}.
Compared with MPI one-sided communication or RDMA Read/Write semantics, it also removes the need to reserve dedicated registered buffers for every communication peer, thereby reducing memory overhead and simplifying memory management.
When the Verbs interface is unavailable on some systems, an MPI-based fallback backend can be used to ensure portability.

%% file: sections/section3_1fig_dsb.tex
\begin{figure}[t]
  \centering
  \includegraphics[width=0.95\linewidth]{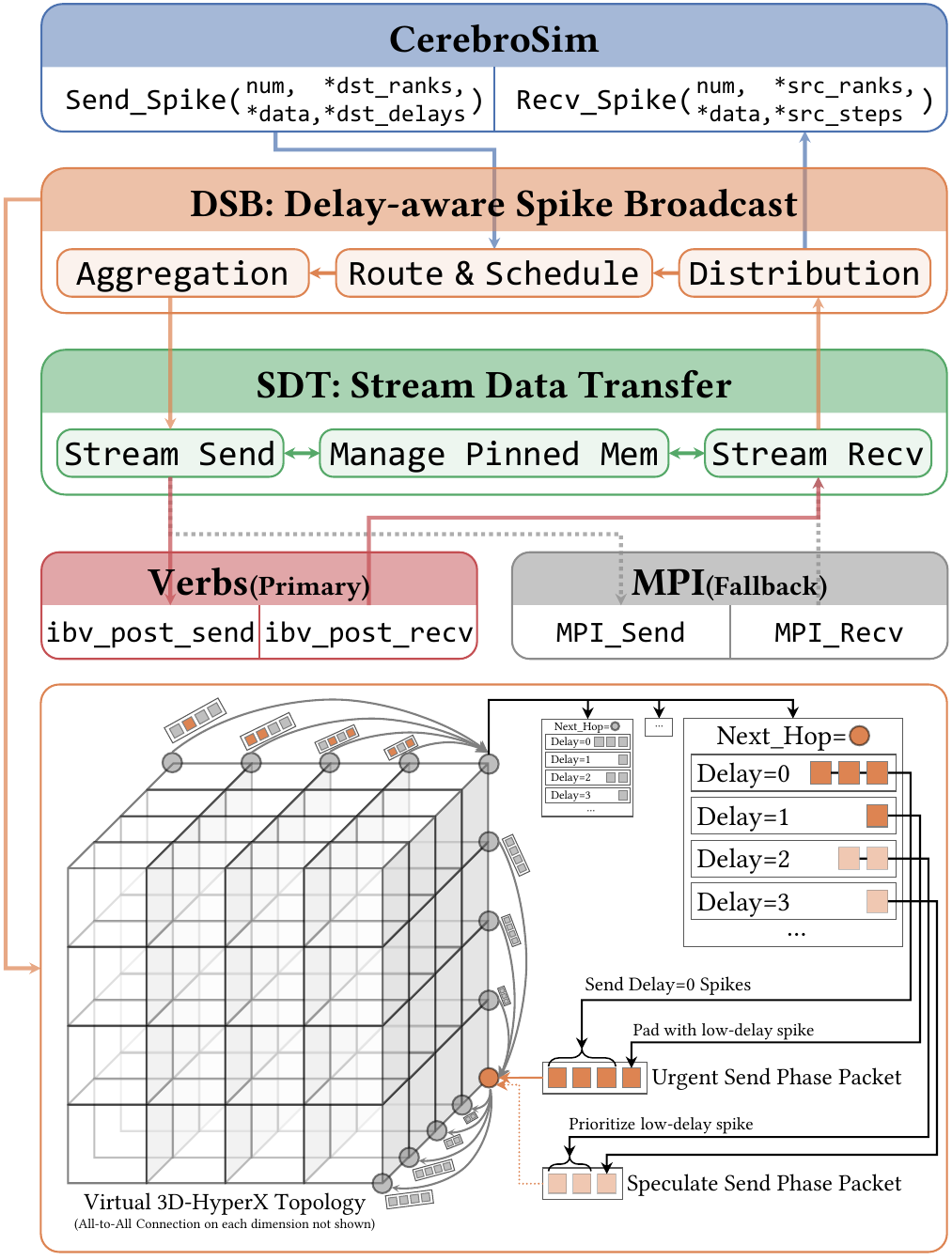}
  \caption{\textbf{DSB: Delay-aware Spike Broadcast.} DSB reduces tiny packets via virtual-topology aggregation and forwarding, overlaps communication with computation by exploiting biological delays, and lowers protocol overhead with Verbs send/recv semantics.}
  \label{fig:dsb}
\end{figure}

%% file: sections/section3_2_rsdc.tex

\input{sections/section3_2fig_rsdc}

Efficient \GlossaryCalibration{spike delivery}{spike delivery} is a critical bottleneck under multithreaded execution, where concurrent updates to shared \GlossaryCalibration{post-synaptic}{post-synaptic} neuron states can introduce race conditions.
Conventional approaches rely on mutex locks or atomic operations to ensure correctness \cite{lu_simulation_2024}, but these mechanisms incur significant synchronization overhead and limit scalability.
To address this challenge, we propose a Race-free Synaptic Dynamics Computation scheme that eliminates the need for locks and atomic primitives while preserving correctness and high performance.

The core idea is to avoid write conflicts during initialization rather than resolving them at runtime, as shown in \Cref{fig:rsdc}.
This is achieved through a combination of data partitioning, deterministic scheduling, and thread-local accumulation.
Instead of allowing multiple threads to update shared synaptic or neuronal states concurrently, we restructure the computation so that each memory location is written by at most one thread at any given time.
A lightweight branch condition is introduced during \GlossaryCalibration{spike delivery}{spike delivery} to filter relevant events without additional data structures.
Importantly, this design introduces no additional memory overhead, as it relies solely on existing \GlossaryCalibration{spike buffers}{spike buffers} and static partition data.
The branch condition replaces more complex routing or synchronization mechanisms, maintaining a minimal and cache-friendly execution path.


During \GlossaryCalibration{spike delivery}{spike delivery}, spikes collected in the \GlossaryCalibration{spike array}{spike buffer} are delivered to their corresponding \GlossaryCalibration{postsynaptic}{post-synaptic} neurons according to locally stored synaptic connections, as shown in \Cref{fig:rsdc}.
The outer loop iterates over active spikes and queries the spike-synapse index to locate the corresponding synapse ranges.
Synapses sharing the same \GlossaryCalibration{presynaptic}{pre-synaptic} neuron are stored contiguously to improve spatial locality.
However, because spiking neural network simulation workloads are intrinsically sparse, accesses to these ranges remain irregular, and each memory access is accompanied by only lightweight computation compared with neuron state updates.
As a result, \GlossaryCalibration{spike delivery}{spike delivery} is primarily memory-bound.

To mitigate this memory bottleneck, we introduce an HBM-assisted software prefetching policy.
Since the \GlossaryCalibration{synapse array}{synapse array} and spike-synapse index are the dominant memory-intensive data structures in \GlossaryCalibration{spike delivery}{spike delivery}, we place them in HBM to exploit its high bandwidth.
We further design two prefetching schemes tailored to the access pattern of \GlossaryCalibration{spike delivery}{spike delivery}.
The Inter-iteration Prefetching scheme prefetches subsequent synapses associated with the \GlossaryCalibration{presynaptic}{pre-synaptic} neuron that generated the current spike into cache while the current synapse is being processed, thereby reducing the latency of later accesses within the same synapse range.
The Inter-loop Prefetching scheme prefetches the first synapse of the next spike while the current spike is being processed, thereby hiding the initial access latency of the next iteration.
Together, these schemes leverage HBM bandwidth and software prefetching to reduce memory stalls and improve \GlossaryCalibration{spike delivery}{spike delivery} efficiency.

%% file: sections/section3_2fig_rsdc.tex
\begin{figure}[t]
  \centering
  \includegraphics[width=1.0\linewidth]{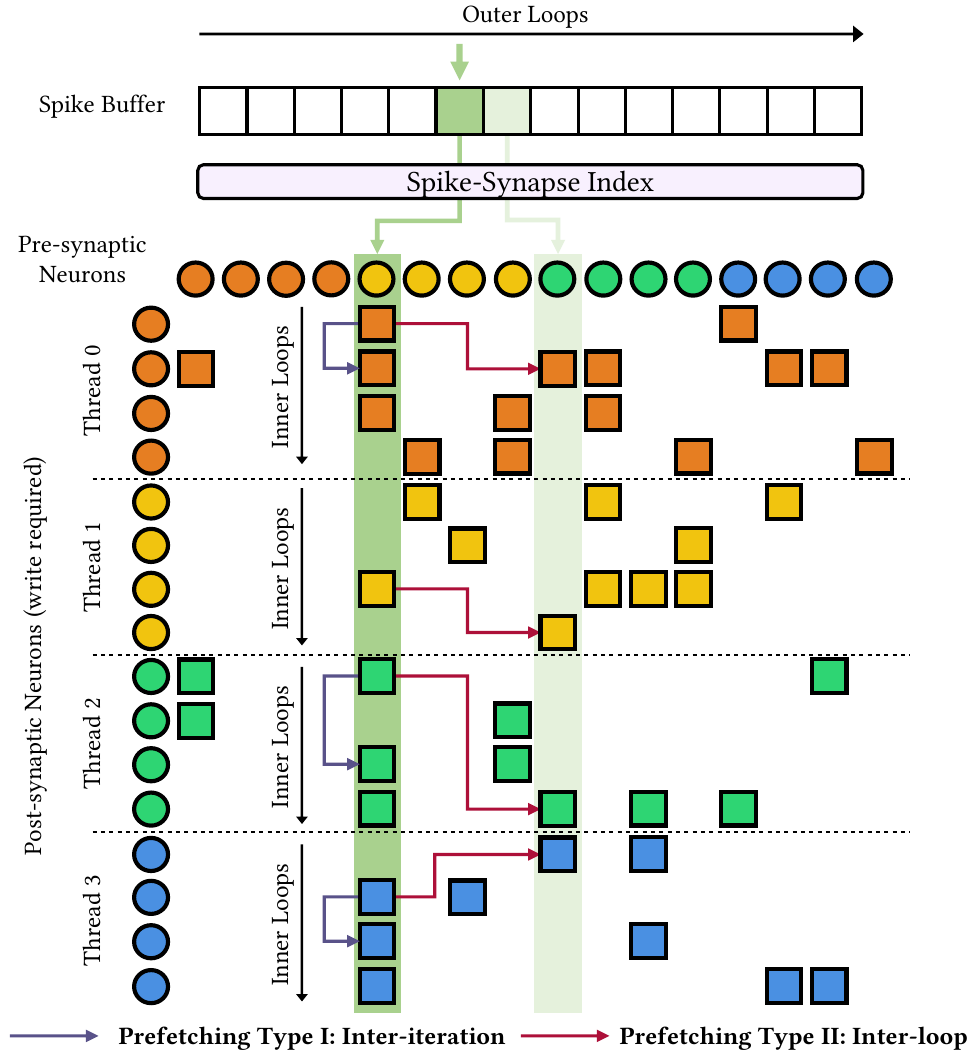}
  \caption{\textbf{RSDC: Race-free Synaptic Dynamics Computation.} Post-synaptic neurons are evenly partitioned across threads; each thread scans the received spike buffer and processes only activated synapses targeting its assigned neurons, enabling race-free neuronal state updates. By further exploiting locality in spike buffer scanning and synapse traversal, two prefetching strategies are introduced to accelerate spike delivery.}
  \label{fig:rsdc}
\end{figure}

%% file: sections/section3_3_saci.tex

\input{sections/section3_3fig_saci}



In whole-brain distributed simulations, inter-voxel connections are extremely sparse.
However, most brain simulators do not fully exploit this property and instead adopt traditional formats to store synapses together with their \GlossaryCalibration{postsynaptic}{post-synaptic} targets.
As a result, such layouts still maintain indices for many \GlossaryCalibration{presynaptic}{pre-synaptic} neuron IDs with no \GlossaryCalibration{postsynaptic}{post-synaptic} targets on the local rank, incurring substantial unnecessary memory overhead.
As shown in \Cref{fig:adj-mat}, Voxel~0 has denser internal connections than connections to other voxels, yet the compressed sparse column example in \Cref{fig:csc} still stores an offset for every \GlossaryCalibration{presynaptic}{pre-synaptic} neuron ID\@.
This index overhead wastes substantial memory and becomes prohibitive at whole-brain scale.



To eliminate this overhead, we use a Synapse-Aware Compressed Index (SACI) to store index entries only for \GlossaryCalibration{presynaptic}{pre-synaptic} neuron IDs with \GlossaryCalibration{postsynaptic}{post-synaptic} targets on the local rank, as shown in \Cref{fig:hashmap}.
By omitting entries for \GlossaryCalibration{presynaptic}{pre-synaptic} neuron IDs with no \GlossaryCalibration{postsynaptic}{post-synaptic} targets on the local rank, this design removes the storage cost of empty index entries and substantially reduces memory usage, especially in whole-brain simulations with sparse inter-voxel connections.

%% file: sections/section3_3fig_saci.tex
\begin{figure}[t]
    \centering
    \subfloat[Adjacency Matrix.\label{fig:adj-mat}]{%
        \includegraphics[width=0.48\linewidth]{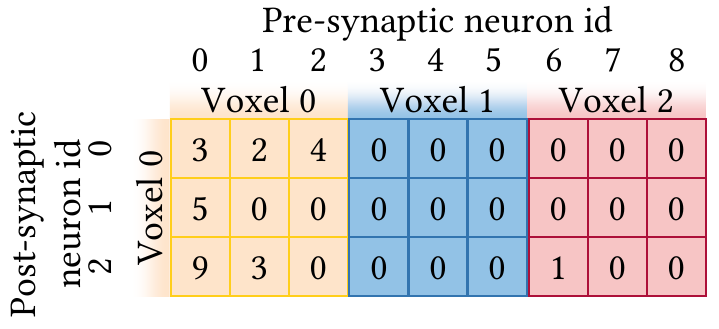}}
    \hfill 
    \subfloat[Compressed Sparse Column.\label{fig:csc}]{%
        \includegraphics[width=0.48\linewidth]{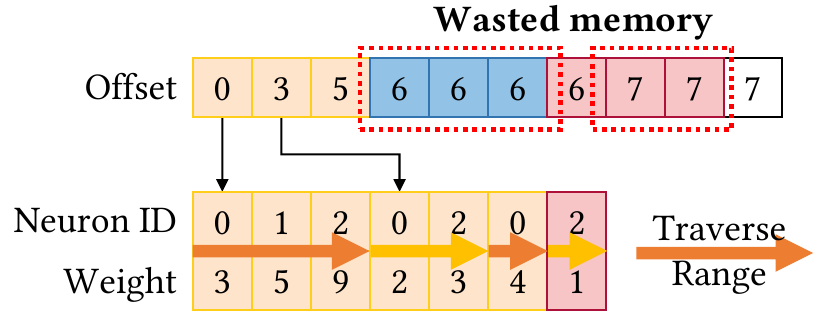}}
    
    
    \vspace{0.2cm} 
    
    \subfloat[SACI: Synapse-Aware Compressed Index (ours).\label{fig:hashmap}]{%
        \includegraphics[width=\linewidth]{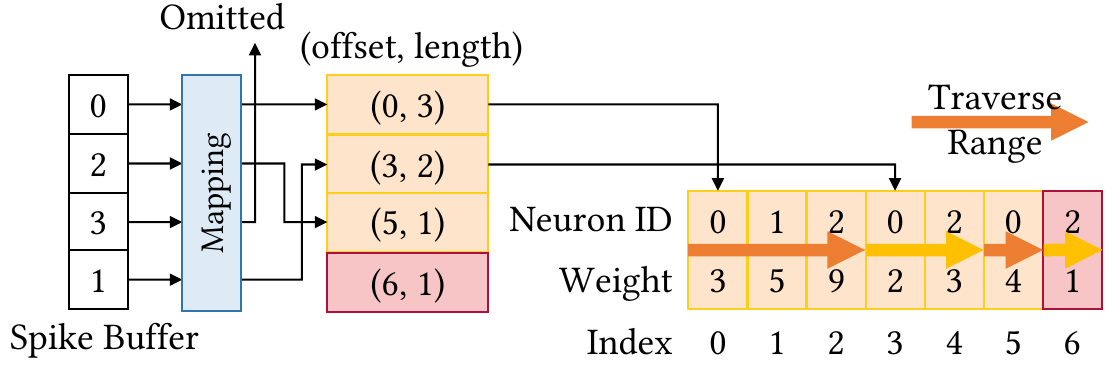}}
    
    \vspace{0.2cm} 
    
    \subfloat[SACI + FlySyn (ours).\label{fig:Regen}]{%
        \includegraphics[width=\linewidth]{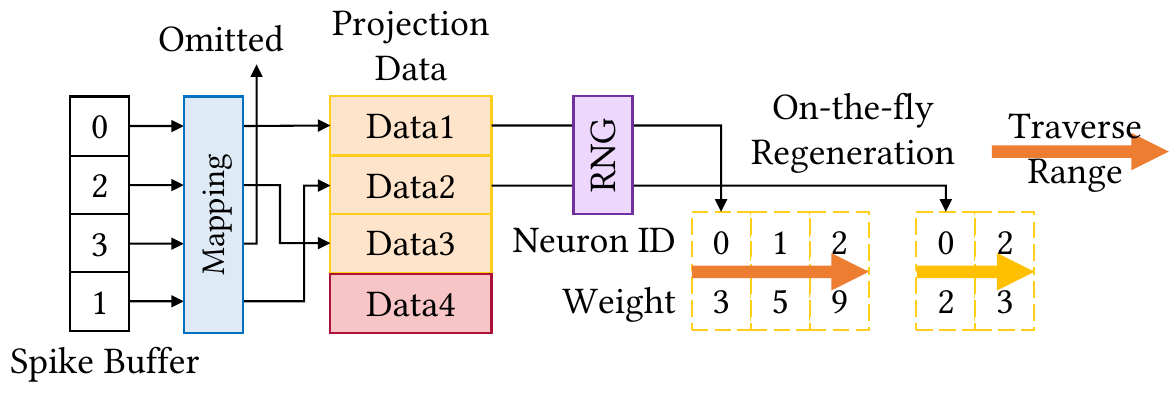}}
    
    \caption{\textbf{Comparison of Synaptic Connection Storage Methods.}
    (a) The adjacency matrix is indexed by pre-synaptic and post-synaptic neuron IDs, with intra-voxel entries denser than inter-voxel entries.
    (b) Compressed sparse column incurs space overhead from maintaining offsets for pre-synaptic neuron IDs with no post-synaptic targets on the local rank.
    (c) SACI compresses index storage by retaining only entries for pre-synaptic neuron IDs with post-synaptic targets on the local rank.
    (d) On-the-fly regeneration removes explicit synapse storage and retains only the index used for regeneration.}
    \label{fig:graph-main-figure}
\end{figure}

%% file: sections/section3_4_flysyn.tex
\input{sections/section3_4fig_smerng}


%
In stochastically defined spiking neural networks, attributes of static synapses—specifically post-synaptic targets and weights—remain invariant throughout the simulation, enabling their deterministic regeneration via pseudo-random generation.
Based on this property, we propose an on-the-fly synapse regeneration scheme that first derives a deterministic seed for synapses due at the current time from \GlossaryCalibration{presynaptic}{pre-synaptic} neuron ID, projection (a structural ensemble of synapses connecting a source population to a target population) ID, rank, thread ID, simulation time, spike time, and global random seed.
The derived seed is then combined with projection data retained in SACI to regenerate synapses. This ensures that spikes emitted by the same \GlossaryCalibration{presynaptic}{pre-synaptic} neuron can regenerate the same synapses on the fly, including their weights, delays, and \GlossaryCalibration{postsynaptic}{post-synaptic} targets. As a result, explicit synapse storage is eliminated, while only projection data used for regeneration is retained in SACI, as shown in \Cref{fig:Regen}.


For static synapses, this reduces the asymptotic memory footprint of the process with rank $i$ from $\mathcal{O}(\alpha_i N + E_i)$ to $\mathcal{O}(\alpha_i N)$. The $\mathcal{O}(\alpha_i N)$ term captures the SACI, while $\mathcal{O}(E_i)$ captures explicit synapse storage, including \GlossaryCalibration{postsynaptic}{post-synaptic} neuron IDs, weights, and delays. Here, $N$ is the total number of neurons in the global simulation, $E_i$ is the number of synapses stored by rank $i$, and $\alpha_i$ is the connectivity coefficient, defined as the fraction of the global neuron population that has \GlossaryCalibration{postsynaptic}{post-synaptic} targets on rank $i$, where a neuron is counted once it has at least one such synaptic connection. At whole-brain scale, $\alpha_i \ll 1$, so on-the-fly regeneration substantially reduces memory consumption.



To meet the substantial throughput requirements of on-the-fly synapse regeneration, we propose \smerng{}, a high-performance random number generator. While designed to be portable across diverse computing environments, \smerng{} is specifically optimized for the ARMv9 architecture on the LineShine supercomputer by harnessing the Scalable Matrix Extension (SME).

We leverage SME to accelerate bit-shuffling, a core mechanism for entropy distribution widely adopted by established generators such as linear congruential generators and Philox \cite{salmonParallelRandomNumbers2011a}. Since multiplication is an inherently effective tool for bit-shuffling, the high-throughput hardware support provided by SME for integer and floating-point matrix outer products offers an ideal primitive for accelerating random number generation. Building upon these hardware primitives, we design a generator architecture, the workflow of which is illustrated in \Cref{fig:sme4rng-workflow}.

Specifically, the SME integer outer product instruction (i.e., \texttt{UMOPA}) interprets each 32-bit lane of the streaming Scalable Vector Extension (SVE) vectors as four 8-bit integers.
This operation performs a $1 \times 4$ by $4 \times 1$ matrix multiplication across corresponding lanes, accumulating the results into the 32-bit elements of a designated ZA tile.
By assigning a fixed multiplier and a state vector to the input streaming vectors, each instruction populates the lower 8 bits of every 32-bit element in the ZA tile with pseudo-random entropy.
Subsequently, four rows of the ZA tile are extracted into SVE vector registers.
We then employ a series of bitwise shifts and OR operations to coalesce these 8-bit segments into a vector of 32-bit random numbers.
Once the ZA tile data is consumed, the state vector $s^{t}$ transitions to $s^{t+1}$ via the xorshift algorithm, preparing the generator for the next iteration.
Taking advantage of the ZA storage's support for up to four concurrent 32-bit tiles, \smerng{} effectively overlaps compute-intensive outer products with SVE-based post-processing.
This hardware-aware design achieves a fully pipelined execution model that successfully hides the latencies of state updates and data extraction.
Beyond architectural efficiency, the generated sequences exhibit high uniformity, as evidenced by the pair-plot.

%% file: sections/section3_4fig_smerng.tex
\begin{figure}[tb]
    \centering
    \includegraphics[width=1.0\linewidth]{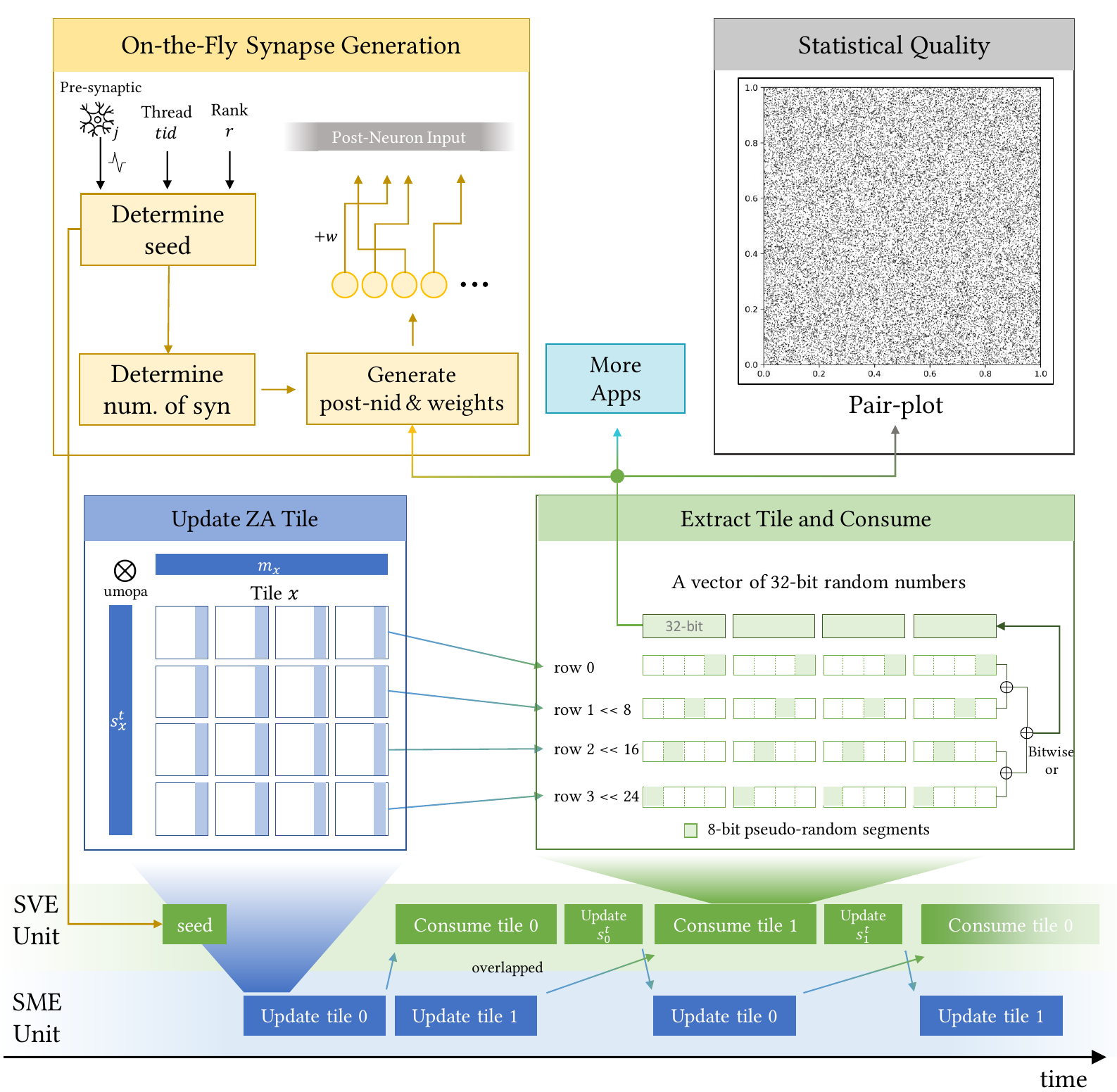}
    \caption{\textbf{On-the-fly Synapse Regeneration via \smerng{}.} The regeneration seed is processed by \smerng{} to load state $s_{x}^{t}$ and multiplier $m_x$. The library populates the 8-bit entropy segments into 32-bit \texttt{ZA} tile elements, which are then extracted and coalesced into 32-bit random numbers for synaptic reconstruction. The pair-plot confirms the robust statistical quality of the generated random values.}
    \label{fig:sme4rng-workflow}
\end{figure}

%% file: sections/section4_setup.tex
\section{Implementation and Experimental Setup}

\subsection{Portable Libraries Design and Implementation}

We extract several key methods from the current simulator into three portable libraries: the DSB Library, the SDT Library, and the \smerng{} Library, so that they can be developed and tested independently and migrated to other application scenarios. For this paper, the reported performance comes from both the integrated execution of these libraries in the \GlossaryCalibration{NERV}{\myname{}} brain simulator and standalone module-level microbenchmarks.


The DSB Library exposes an application-facing abstraction for Delay-aware Spike Broadcast, where the application specifies destination ranks, per-target delays, and message sizes.
Inside the DSB Library, route construction, next-hop aggregation, delay-aware scheduling, and exact-step \GlossaryCalibration{delivery}{propagation} are encapsulated behind a stable interface, removing complex communication logic from the \GlossaryCalibration{NERV}{\myname{}} code base.
This packaging allows the DSB Library to evolve independently and to migrate to other distributed brain simulators without reimplementing the same low-level mechanisms in each code base.

The SDT Library provides a unified streaming transport interface for the DSB Library and other applications that require efficient streaming data movement.
Its high-performance path is built primarily on Verbs send/recv semantics together with pre-posted receives and pooled registered buffers, reducing remote address exchange, repeated memory registration, and unnecessary data copies.
When Verbs is unavailable on a target system, the SDT Library can use its MPI-based fallback backend for functional execution and correctness validation, or can achieve high performance by adding a backend for the target system's interconnect without requiring repeated modifications to the DSB Library, \GlossaryCalibration{NERV}{\myname{}}, or other upper-layer software.


The \smerng{} Library features a dual-tier API architecture to accommodate diverse integration requirements.
The high-level API offers streamlined and intuitive functions for random number generation, specifically designed to accelerate throughput-intensive applications requiring massive on-the-fly stochasticity.
In contrast, the low-level API grants users direct access to random sequences within vector registers.
By exposing these low-level primitives, the library facilitates seamless co-optimization with vectorized application logic, allowing fine-grained control over data movement and functional unit utilization.

\input{sections/section4_1fig_rasters}

\input{sections/section4_2fig_scalability}

\subsection{Whole-Brain Simulation Case}

We present a large-scale whole-brain simulation of spiking neuronal networks \cite{gerstner_single_nodate} at human-brain scale, derived from diffusion tensor imaging data \cite{gupta_review_2008} and biologically constrained connectivity, as shown in \Cref{fig:overview}.
The system is designed to reproduce resting-state dynamics in both epilepsy \cite{stafstrom_seizures_2015} and normal conditions.
The implementation scales to 86 billion neurons and runs on 18{,}432 compute nodes, demonstrating strong and weak scaling at extreme scale.
These results demonstrate the feasibility of constructing a biologically constrained human brain model on a leadership-class supercomputer, providing a foundation for studies of neuronal dynamics and brain disorders.

In this work, we transform MRI-derived data into a neuronal network architecture according to the following design principles.
First, the number of neurons assigned to each voxel is assumed to be proportional to the local gray matter volume estimated from longitudinal relaxation time weighted MRI\cite{gaeta_t1_2024}.
Based on this mapping, neurons are distributed across brain regions as follows: the cortex (18{,}432 voxels) and subcortical structures---including the subcortex, brainstem, and cerebellum (1{,}536 voxels)---comprising approximately 16 billion and 70 billion neurons, respectively.
Second, structural connectivity between voxel pairs is parameterized using a row-normalized, voxel-wise diffusion-weighted imaging connectivity matrix \cite{chilla_diffusion_2015}.
Specifically, each matrix element is normalized by the sum of all elements in its corresponding row, yielding a probabilistic interpretation of outgoing connections from each source voxel.
This formulation captures white matter micro-structural organization and biophysical constraints inferred from local diffusion properties.

Within each voxel, neurons are modeled using a microcircuit model based on leaky integrate-and-fire neurons \cite{caceres_analysis_2011}, driven by \GlossaryCalibration{post-synaptic}{post-synaptic} currents as inputs and supplemented with independent Poisson processes to capture non-synaptic background activity.
Synaptic connectivity between neurons is defined as static, with transmission delays and synaptic weights drawn from normal distributions.
Synaptic plasticity \cite{citri_synaptic_2008} is not incorporated in the current model because the biological evidence remains insufficient to support reliable large-scale parameterization of plasticity mechanisms across the whole brain.


Based on 5,638 neurons selected from four voxels of the Primary Somatosensory Cortex \cite{borich_understanding_2015}, the two raster plots in \Cref{fig:rasters} show neuronal activity under epilepsy (a) and normal (b) conditions, with neurons grouped by layer and type (excitatory in blue, inhibitory in red).
In the epilepsy condition (left), several abnormalities are visible.
Excitatory populations, particularly in L2/3E and L4E, exhibit more irregular and clustered firing, with noticeable fluctuations in spike density over time.
This pattern suggests disrupted input integration and unstable recurrent excitation.
At the same time, inhibitory populations maintain relatively consistent firing bands, but their activity appears less tightly coupled to excitatory fluctuations, indicating a breakdown in excitation-inhibition (E/I) coordination.
The deeper layer L6E remains sparse but shows slightly increased variability.
Overall, the epilepsy network exhibits higher temporal variability and reduced stability, consistent with impaired synaptic transmission and network dysregulation.
In contrast, the normal condition (right) shows a more homogeneous and stable firing pattern across all layers.
Excitatory neurons, especially in L4E, display uniform spike distributions over time, reflecting stable recurrent dynamics and consistent external drive.
Inhibitory populations form well-defined, steady bands, indicating inhibitory control and tight coupling with excitatory activity.
The balance between excitation and inhibition is preserved, resulting in an asynchronous yet stable regime without excessive clustering or silence.

%% file: sections/section4_1fig_rasters.tex
\begin{figure}[b] 
	\centering
    \subfloat[Raster Plot of Epilepsy.\label{fig:raster_ad}]{%
        \includegraphics[width=0.24\textwidth]{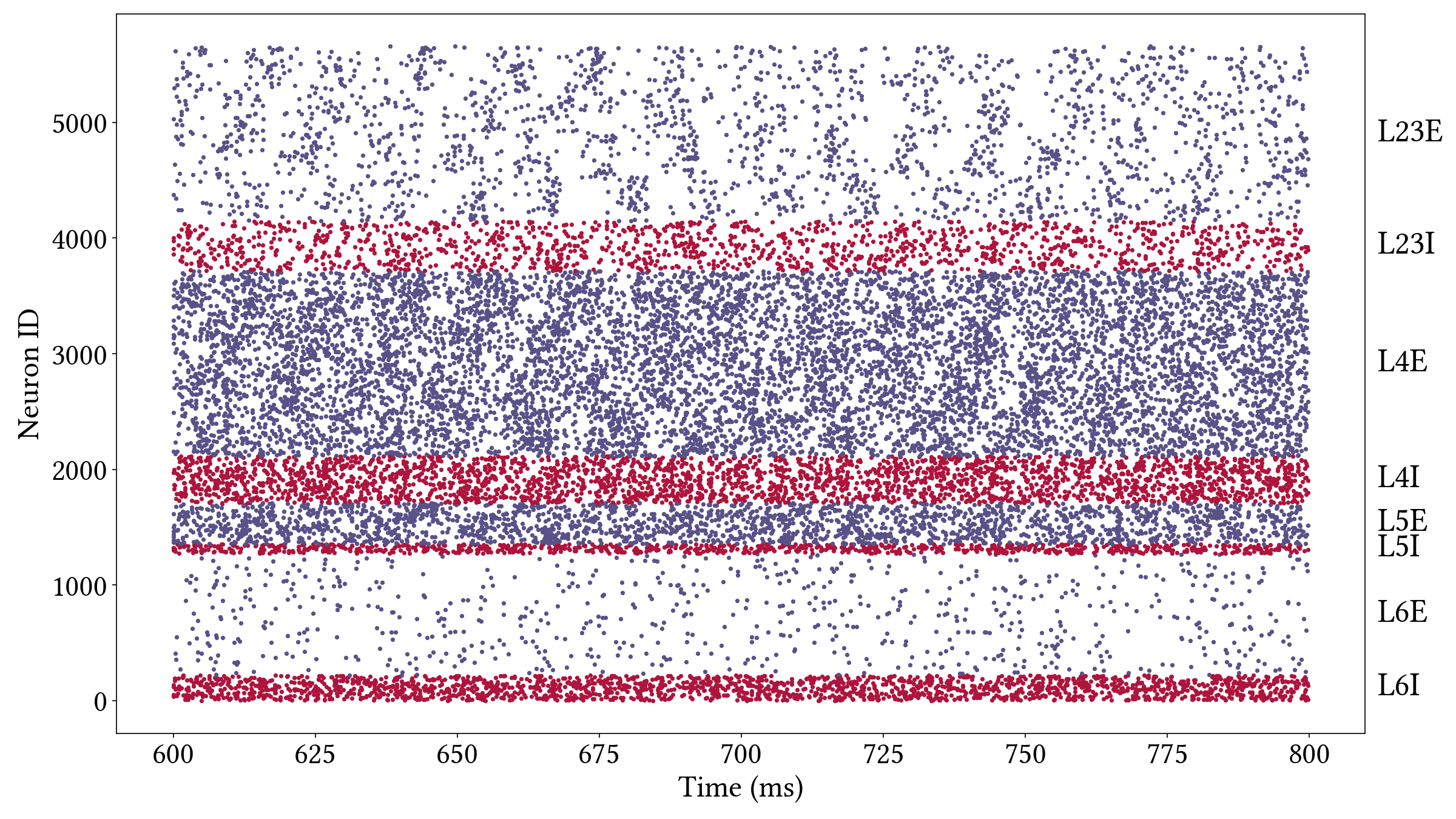}}
    \subfloat[Raster Plot of Normal.\label{fig:raster_normal}]{%
        \includegraphics[width=0.24\textwidth]{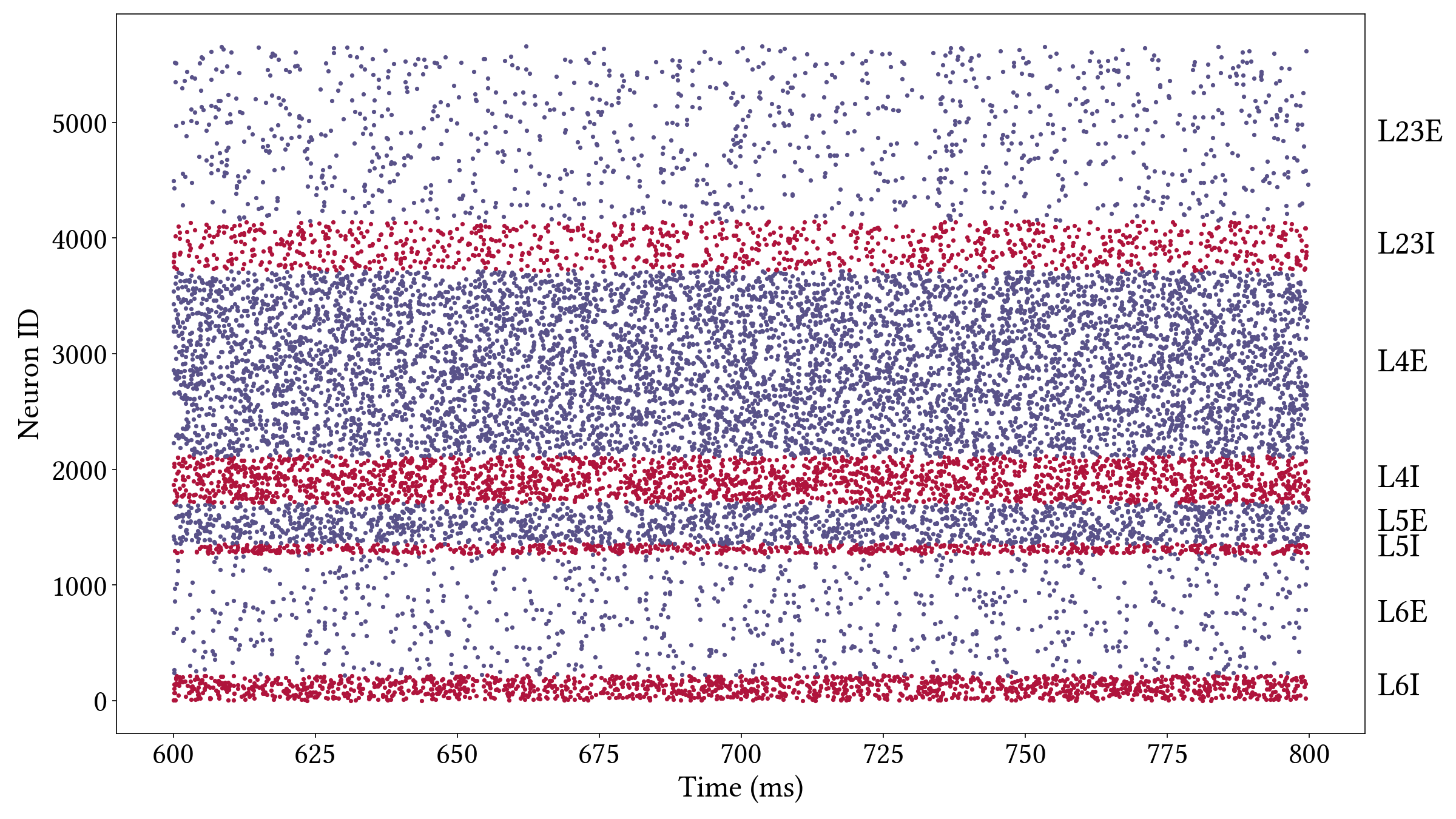}}
    \caption{\textbf{Raster Plots of the Simulation Results.} The horizontal axis represents time, and the vertical axis represents neurons. Each dot or tick mark indicates the moment at which a neuron fires a spike.}
    \label{fig:rasters}
\end{figure}

%% file: sections/section4_2fig_scalability.tex
\begin{figure*}[th]
  \centering
  \subfloat[Weak Scalability.\label{fig:weak_scalability}]{%
  \includegraphics[width=0.51\textwidth,keepaspectratio]{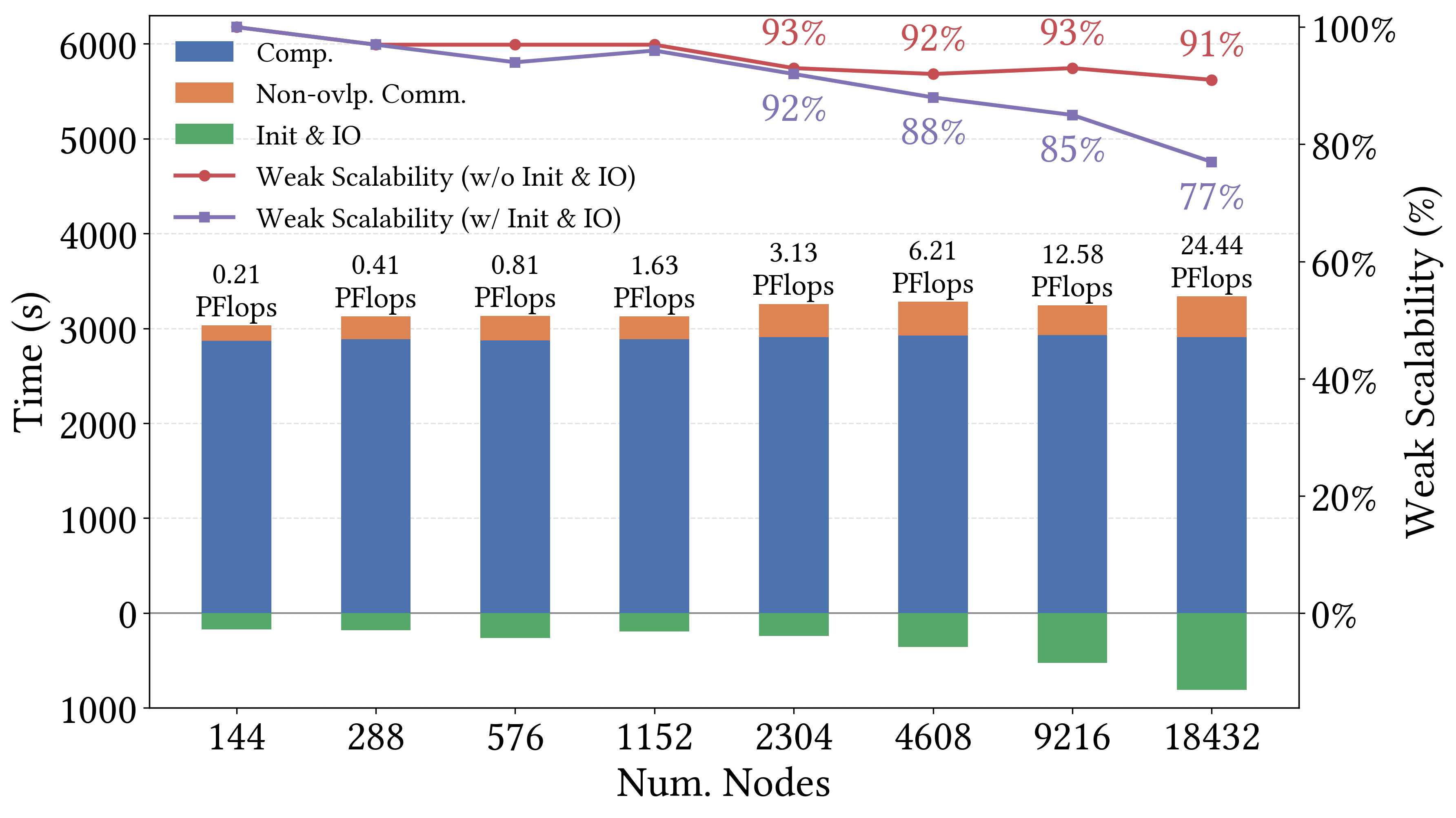}}\hfill
  \subfloat[Strong Scalability.\label{fig:strong_scalability}]{%
  \includegraphics[width=0.48\textwidth,keepaspectratio]{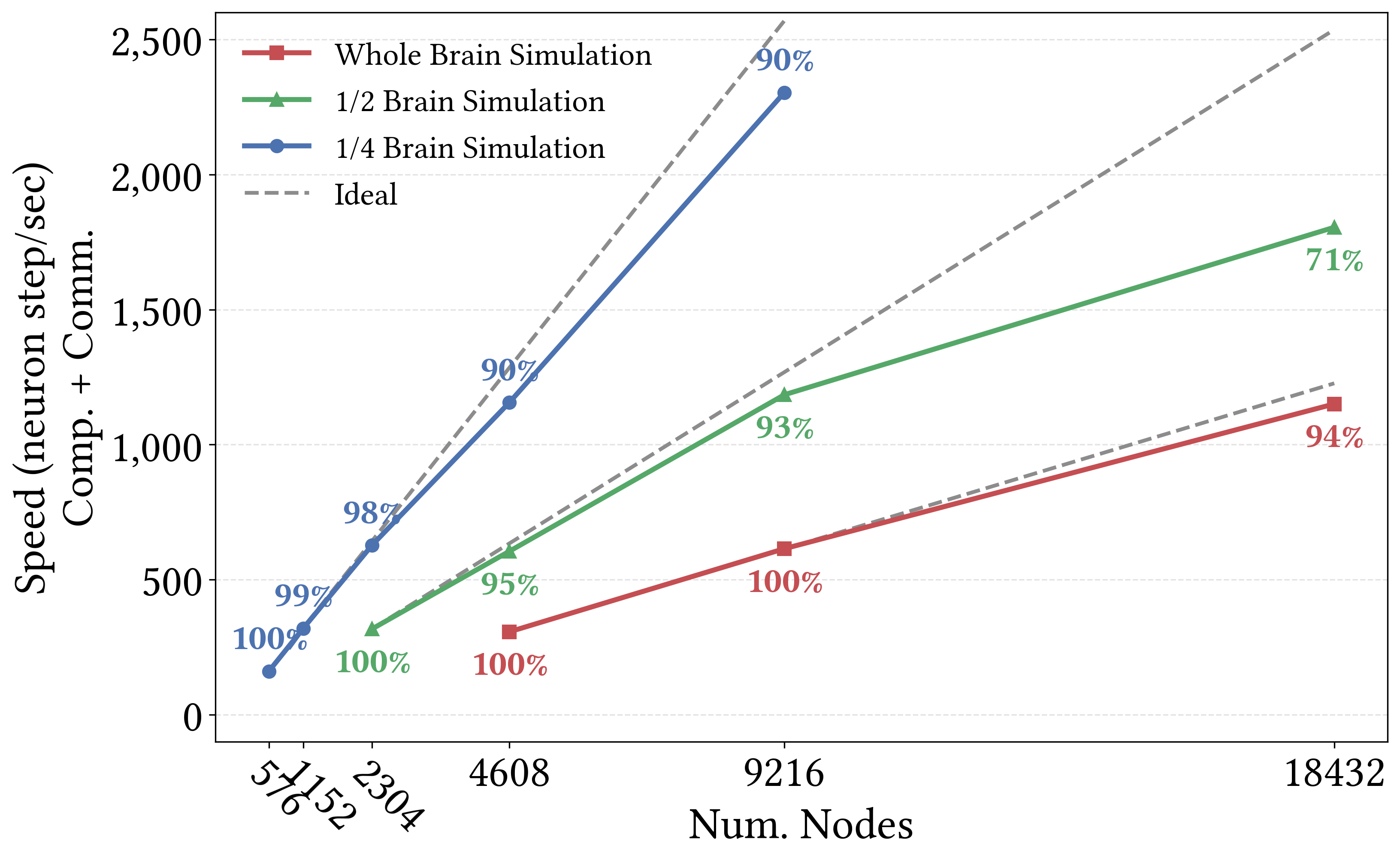}}
  \caption{\textbf{\myname{} Scalability Evaluation.} \myname{} maintains high scalability in both weak- and strong-scaling evaluations; for the whole-brain case, it achieves 91\% weak-scaling efficiency in the core simulation phase and 94\% strong-scaling efficiency, while sustained performance reaches 24.44~PFlop/s at 18{,}432 nodes.}
  \label{fig:scalability_figures}
\end{figure*}

%% file: sections/section5_evaluation.tex
\section{Performance Evaluation}

\subsection{Scalability}

As shown in \Cref{fig:weak_scalability}, the core simulation stage maintains high weak scalability from 144 to 18{,}432 nodes.
The computation time remains nearly constant, staying in a narrow range from 2867 to 2930 seconds, while the non-overlapped communication time increases only moderately.
The sustained performance correspondingly increases from 0.21~PFlop/s at 144 nodes to 24.44~PFlop/s at 18{,}432 nodes, indicating near-proportional growth in aggregate throughput as the system scales out.
When initialization and finalization are excluded, the weak-scaling efficiency still reaches 91\% at 18{,}432 nodes.
When these stages are included in the end-to-end measurement, the weak-scaling efficiency drops to 77\%.
This gap is mainly caused by the relatively fixed overhead of initialization and finalization rather than by substantial degradation in the scalability of the main iterative simulation phase.


As shown in \Cref{fig:strong_scalability}, we evaluate strong scalability using whole-brain, half-brain, and quarter-brain cases.
The whole-brain simulation has the largest memory footprint and therefore requires at least 4{,}608 nodes to run, whereas smaller cases can be tested on fewer nodes.
This multi-case design allows us to assess strong scalability both for the full application at near-production scale and across a much wider node range.
For the quarter-brain case, scaling from 576 to 9{,}216 nodes achieves 90\% scalability, demonstrating near-linear and robust scaling over this wide node range.
For the half-brain case, scaling from 2{,}304 to 9{,}216 nodes sustains 93\% scalability, but extending further to 18{,}432 nodes reduces scalability to 71\% because communication can no longer be well hidden by computation.
For the whole-brain case, scaling from 4{,}608 to 18{,}432 nodes still maintains 94\% strong scalability.

\subsection{Effectiveness of the Optimizations}

This subsection uses targeted ablation experiments to quantify the contribution of each key optimization.

\subsubsection{Delay-Aware Spike Broadcast}
\input{sections/section5_1_dsb}

\subsubsection{Race-Free Synaptic Dynamics Computation}
\input{sections/section5_2_rsdc}

\input{sections/section5_2fig_rsdc}

\subsubsection{Sparse Synapse Storage Compression}
\input{sections/section5_3_3sc}

%% file: sections/section5_1_dsb.tex
\input{sections/section5_1fig_dsb}

As shown in \Cref{fig:bsb_nerv_cmp}, we compare the performance of \GlossaryCalibration{NERV}{\myname{}} with the DSB implementation and the MPI-collective implementation.
DSB fully exploits the delay-tolerant nature of \GlossaryCalibration{spike communication}{spike propagation} in brain simulation to overlap communication with computation, and the small remaining non-overlapped portion arises from the limited set of messages that must be processed in the current communication step.
DSB also provides an additional benefit by significantly reducing spike-filtering overhead.
Because DSB performs targeted broadcasts, each process receives primarily the data that it actually needs to consume locally.
In contrast, the MPI collective implementation primarily uses \texttt{MPI\_Allgatherv}, which forces all processes to receive and filter a large volume of irrelevant messages (\texttt{MPI\_Alltoallv} is impractical at this scale because it would create too many point-to-point connections).
This joint reduction in communication and filtering overhead yields 59\%--262\% end-to-end application performance gains over MPI collectives and is therefore one of the key reasons why \GlossaryCalibration{NERV}{\myname{}} scales well.

\Cref{fig:bsb_cmp_144,fig:bsb_cmp_576,fig:bsb_cmp_2304} show the performance impact of different network-topology configurations under DSB.
We evaluate these configurations with a microbenchmark that reproduces the communication characteristics of \GlossaryCalibration{NERV}{\myname{}} at the corresponding scale, including HyperX virtual topologies in 2D, 3D, and 4D, a Dragonfly virtual topology, and a fallback version that uses the MPI backend instead of the SDT Verbs backend.
Among the evaluated virtual topologies, the 3D HyperX configurations are the best choices because they better balance virtual path length against per-hop traffic load, improving performance by 10\%--94\% over the other virtual topologies.

At the same time, the fallback mode that uses MPI as the SDT communication backend, although 19\%--35\% slower than the best high-performance Verbs backend, remains competitive, indicating that DSB does not depend entirely on Verbs support to be effective.

%% file: sections/section5_1fig_dsb.tex
\begin{figure*}[t]
  \centering
  \subfloat[DSB vs. MPI Collectives.\label{fig:bsb_nerv_cmp}]{%
  \includegraphics[height=0.14\textheight,keepaspectratio]{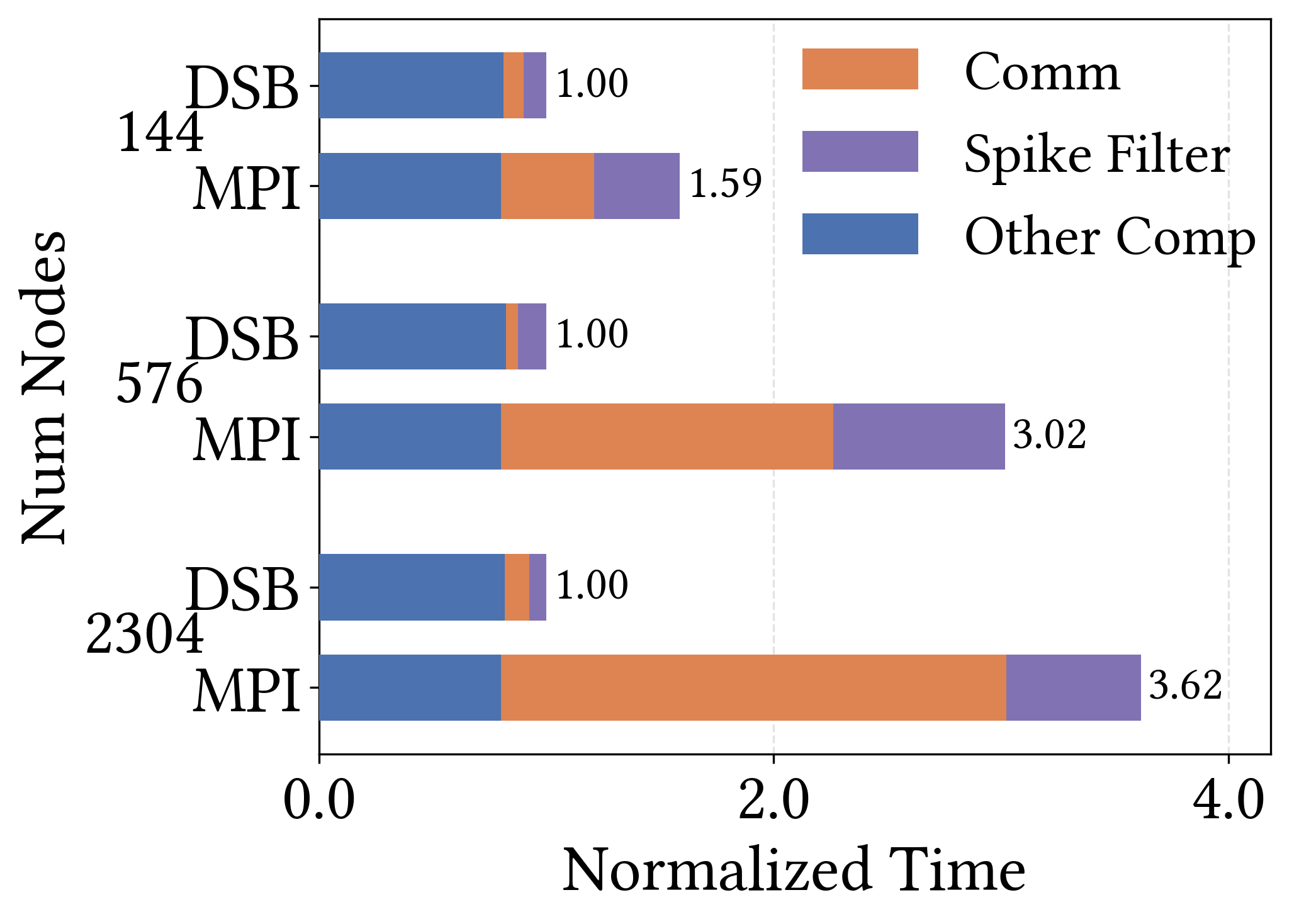}}\hfill
  \subfloat[144-Node Configuration.\label{fig:bsb_cmp_144}]{%
  \includegraphics[height=0.14\textheight,keepaspectratio]{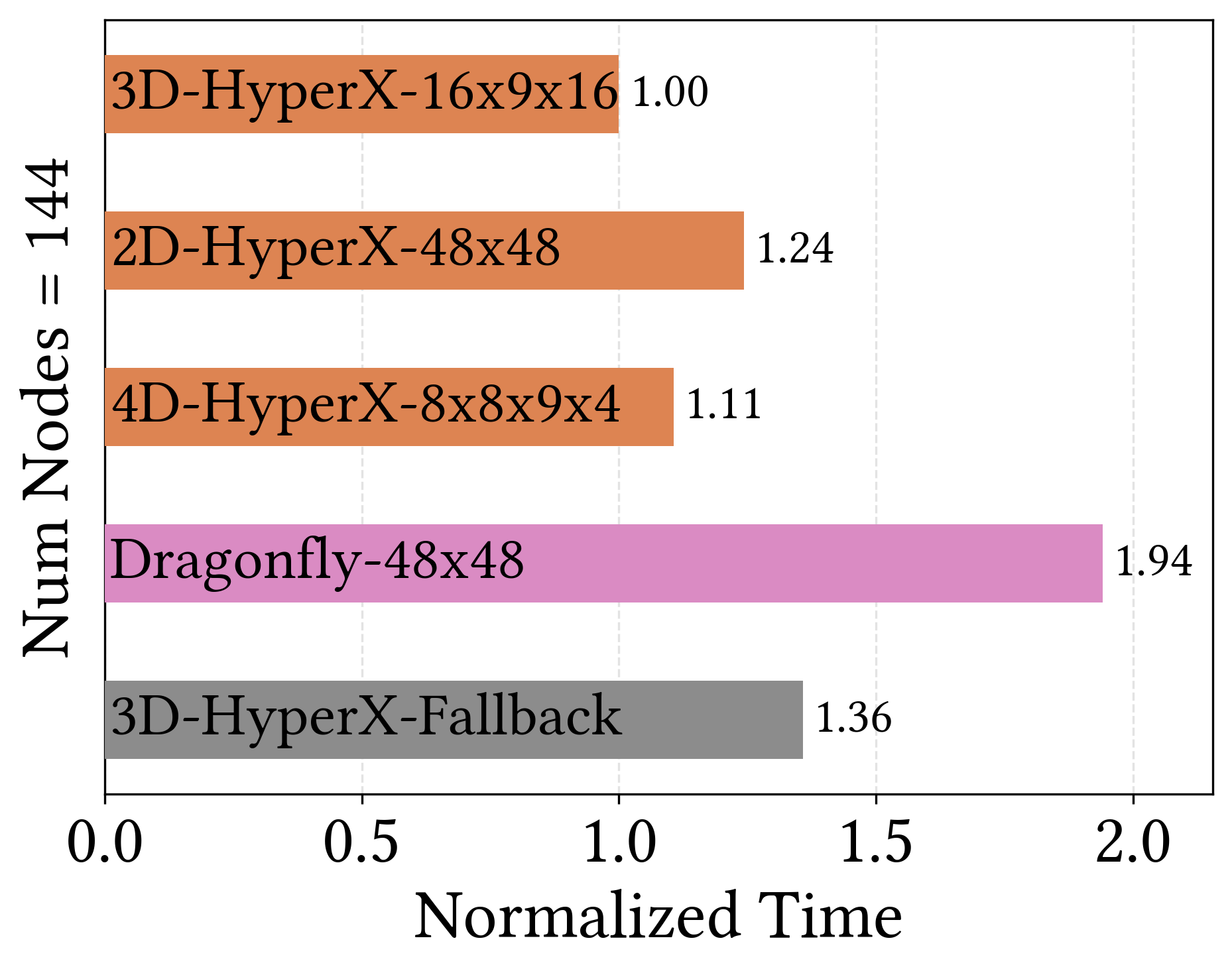}}\hfill
  \subfloat[576-Node Configuration.\label{fig:bsb_cmp_576}]{%
  \includegraphics[height=0.14\textheight,keepaspectratio]{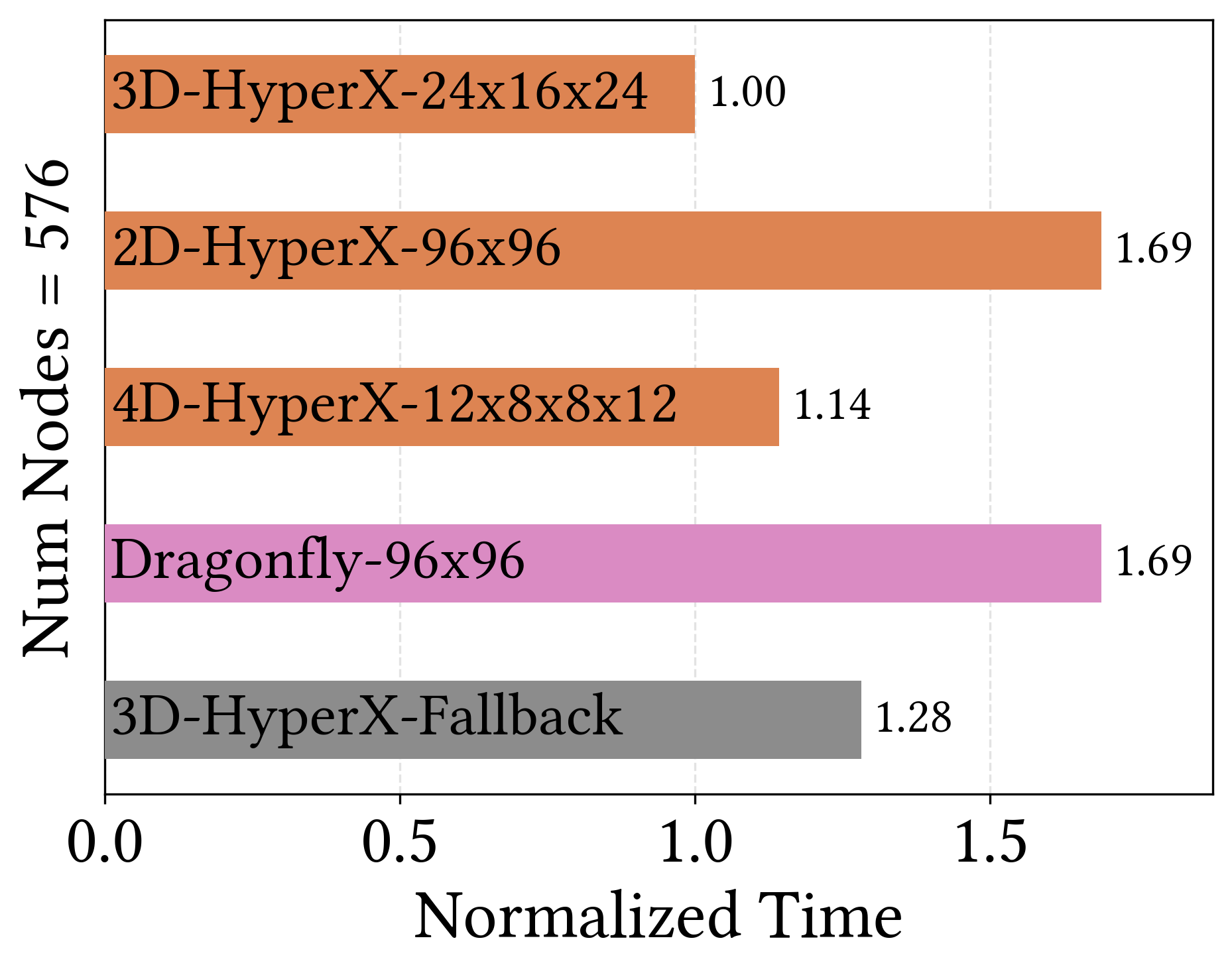}}\hfill
  \subfloat[2{,}304-Node Configuration.\label{fig:bsb_cmp_2304}]{%
  \includegraphics[height=0.14\textheight,keepaspectratio]{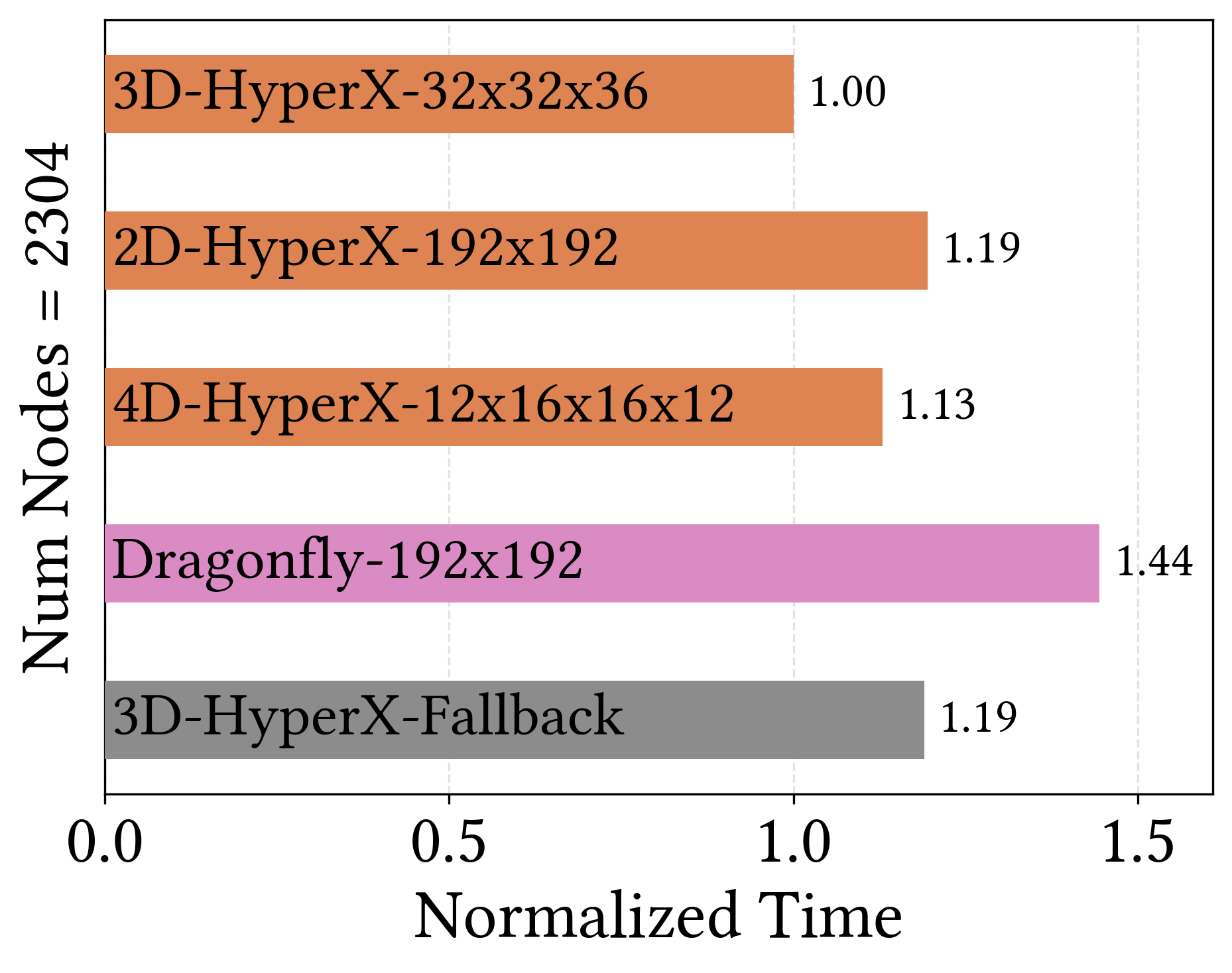}}
  \caption{\textbf{DSB Performance Evaluation.} DSB hides most communication behind computation and reduces spike-filtering overhead compared with MPI collectives; across 144, 576, and 2{,}304 nodes, 3D HyperX is consistently the best or near-best virtual topology, while the MPI fallback backend remains competitive.}
  \label{fig:dsb_figures}
\end{figure*}

%% file: sections/section5_2_rsdc.tex
To quantify the impact of \GlossaryCalibration{spike delivery}{spike delivery} optimization, we conduct an ablation study comparing the baseline implementation with an optimized multithreaded scheduling strategy across three system scales: 144, 288, and 576 nodes.
As shown in \Cref{fig:spk_omp}, the total execution time is decomposed into two components: \MyComment{\GlossaryCalibration{Spike Delivery}{Spike Delivery} and Other}{HH：大小写？} (computation, synchronization, and miscellaneous overheads).

The optimized scheduling method consistently reduces total runtime across all scales.
These results demonstrate that the optimization scales effectively with increasing system size.
The ``Other'' component also shows a modest reduction (e.g., from 66.1s to 45.9s at 144 nodes), suggesting that improved scheduling indirectly enhances load balance and reduces synchronization overhead.
However, this improvement is secondary compared to the gains in \GlossaryCalibration{spike delivery}{spike delivery}.


\Cref{fig:hbm_policy_eval} evaluates how HBM data placement and software prefetching policies affect the \GlossaryCalibration{spike delivery}{spike delivery} stage.
Placing either the spike-synapse index or the \GlossaryCalibration{Synapse Array}{synapse array} in HBM improves performance, while placing both structures in HBM yields the largest speedup; therefore, all prefetching policies are evaluated with this data placement. For Inter-iteration Prefetching, all tested prefetching configurations improve performance, with a prefetch stride of 1 achieving the best result. This is enabled by the \GlossaryCalibration{Synapse Array}{synapse array} layout, which stores synapses that are likely to be accessed together contiguously. Larger prefetch strides, however, degrade performance due to the spatiotemporal sparsity of spiking neural network simulation workloads: even when related synapses are colocated, only a small subset may be active at a given time step, causing large-stride prefetching to fetch unnecessary data. Based on the optimal Inter-iteration Prefetching configuration, we further evaluate Inter-loop Prefetching, which still provides a modest speedup despite crossing the inner loop and confirms the effectiveness of the proposed prefetching policy.

%% file: sections/section5_2fig_rsdc.tex
\begin{figure*}[t] 
    \centering
    
    \begin{minipage}{0.66\textwidth}
        \centering
        \subfloat[Race-free Spike Delivery.\label{fig:spk_omp}]{%
            \includegraphics[width=0.48\textwidth]{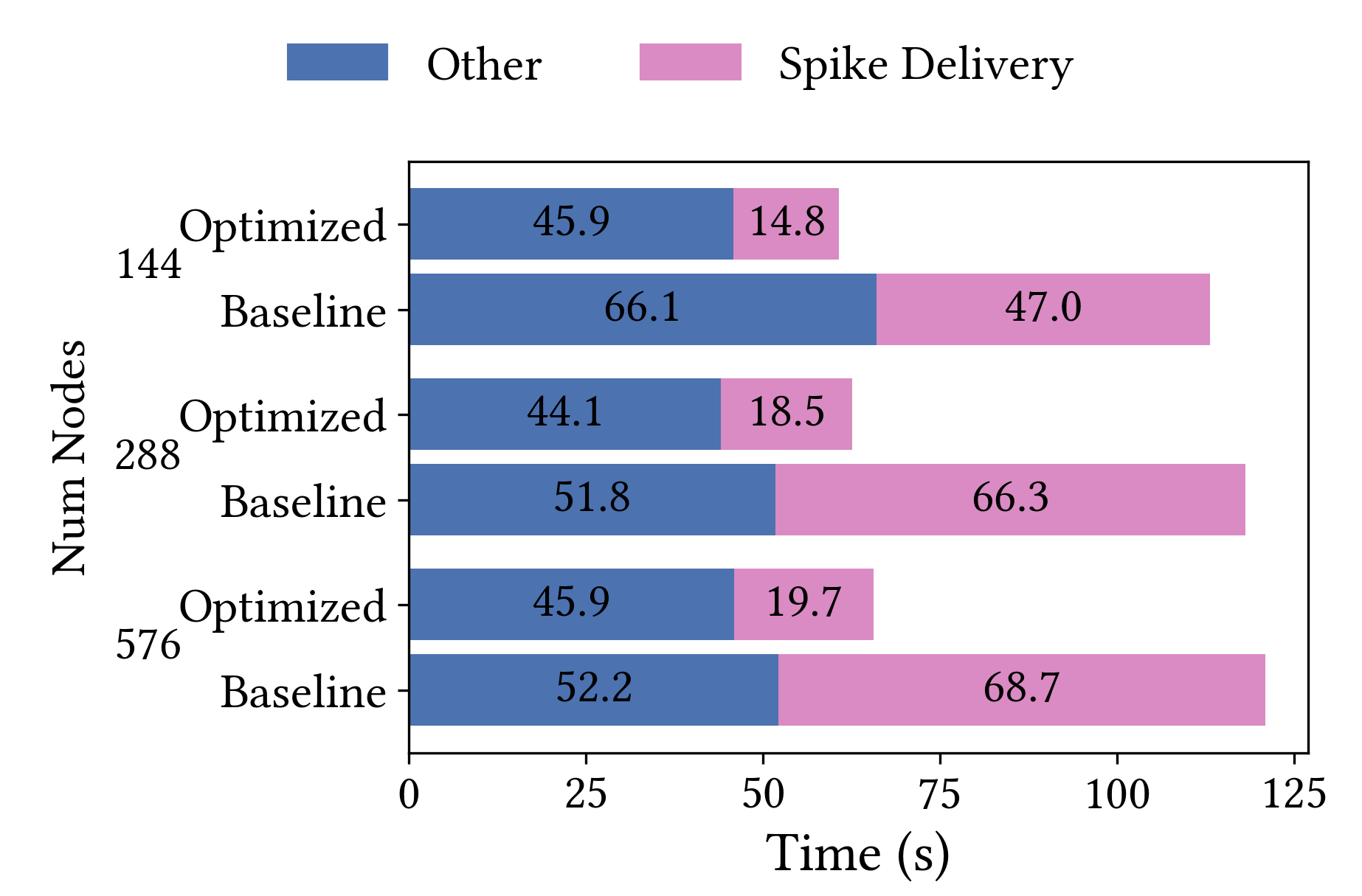}}
        \hfill
        \subfloat[HBM Prefetching Policy Evaluation.\label{fig:hbm_policy_eval}]{%
            \includegraphics[width=0.48\textwidth]{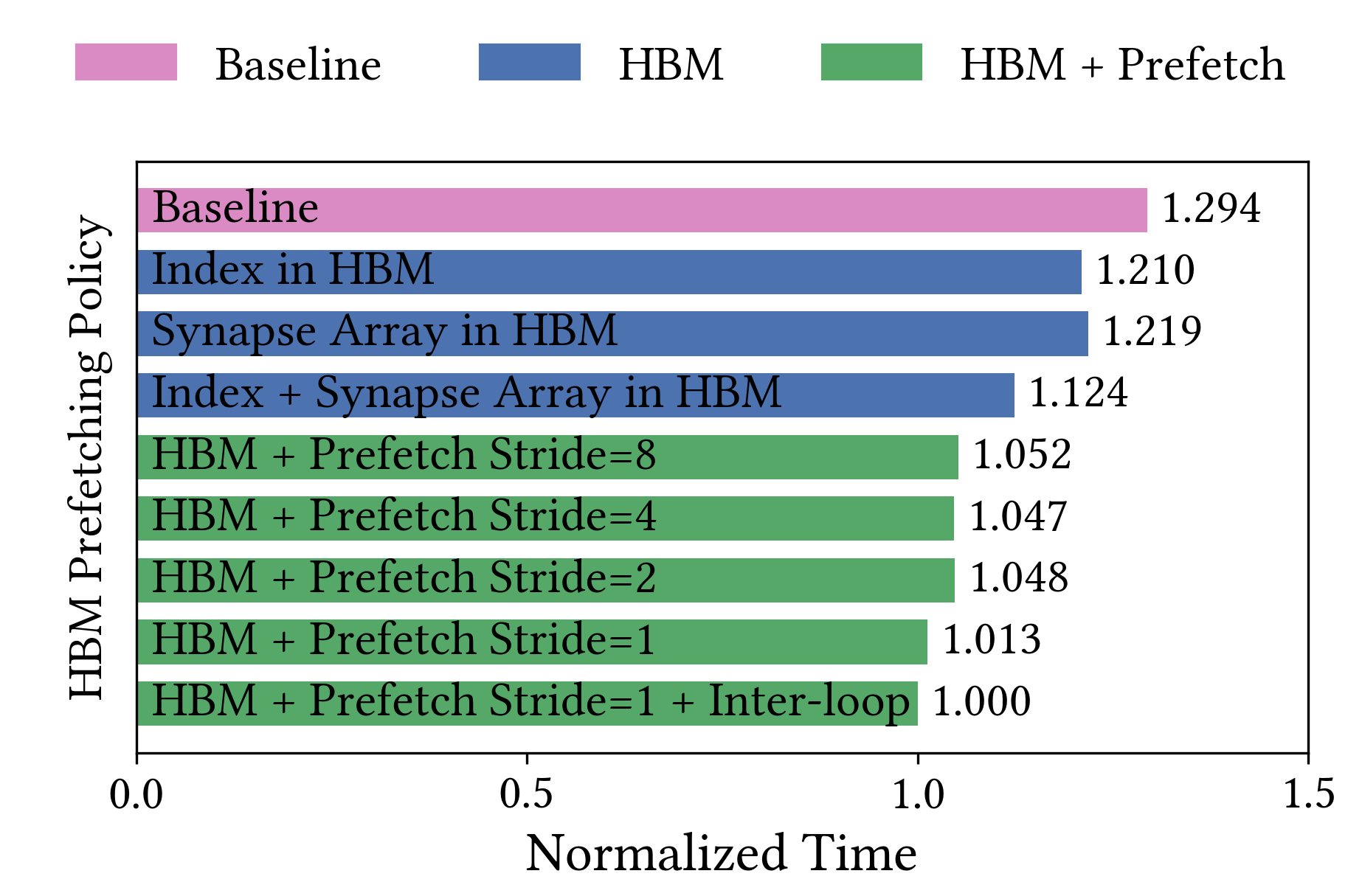}}

        \caption{\textbf{Race-free Synaptic Dynamics Computation Evaluation.}
        (a) RSDC substantially accelerates spike delivery by eliminating costly locks and atomic operations.
        (b) Placing spike-synapse index and synapse arrays in HBM improves memory-access performance, while Inter-iteration and Inter-loop prefetching provide further optimization.}

        \label{fig:rsdc_evaluation}
    \end{minipage}
    \hfill 
    \begin{minipage}{0.32\textwidth}
        \centering
        \includegraphics[width=\textwidth]{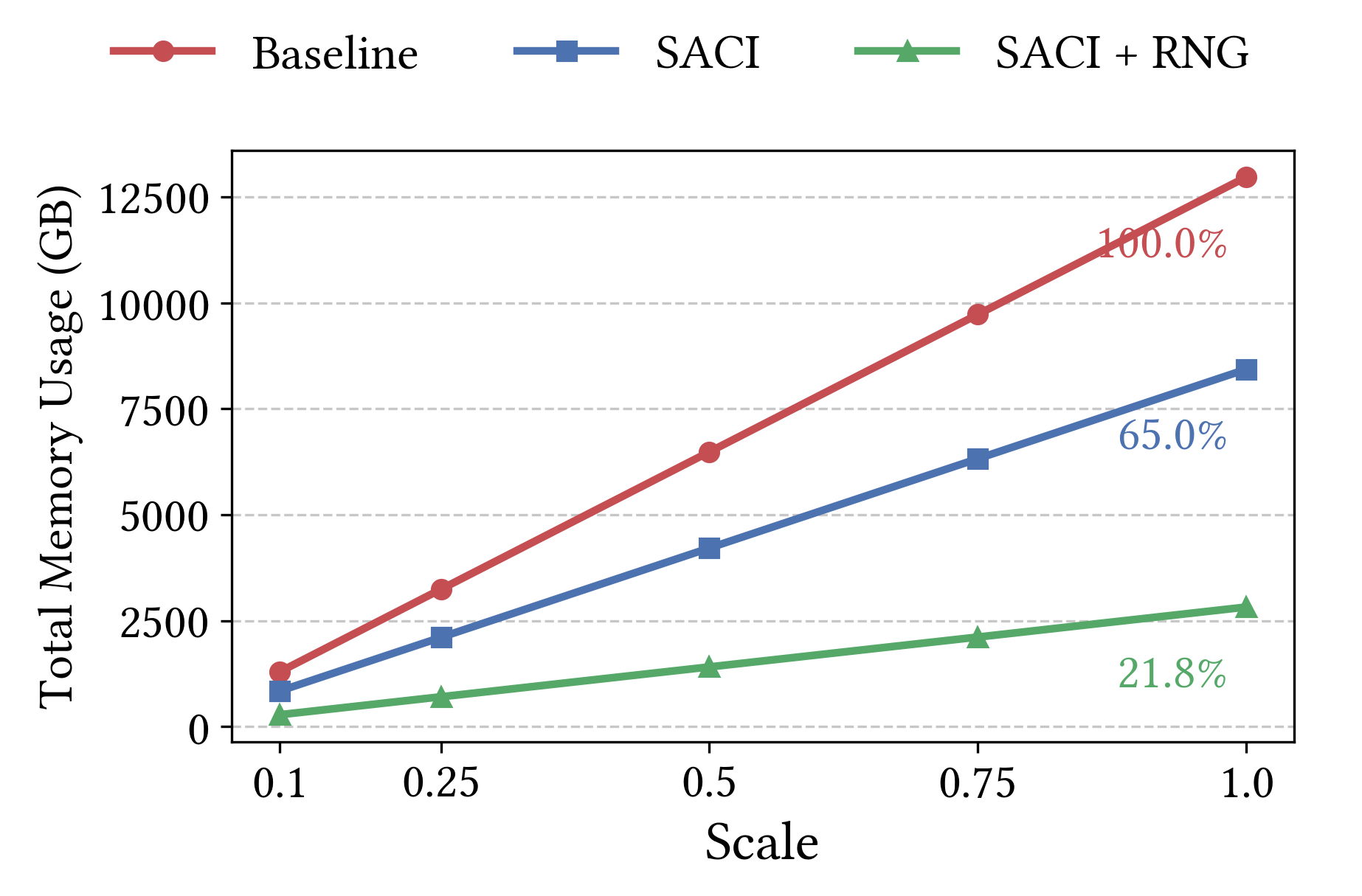}
\caption{\textbf{Memory Compression Evaluation.}
Results on 144 nodes covering 0.78\% of brain regions show the memory reduction achieved by SACI and the further savings from on-the-fly regeneration.}

        \label{fig:mem_compression}
    \end{minipage}
    
\end{figure*}

%% file: sections/section5_3_3sc.tex


We evaluate the Synapse-Aware Compressed
Index scheme and the random number generator (RNG)-based on-the-fly regeneration scheme on 144 nodes over approximately 0.78\% of the whole-brain regions, with scale controlling the number of neurons per region. As shown in \Cref{fig:mem_compression}, the SACI scheme reduces memory usage to 65.0\% of the baseline by storing index entries only for nonzero entries. The RNG-based scheme further reduces memory usage to 21.8\% by eliminating explicit synapse storage and retaining only the projection data needed for RNG-based regeneration. The connectivity coefficient $\alpha$ is mainly affected by the number of voxels, because biological inter-voxel connections are far sparser than intra-voxel connections. Therefore, increasing scale changes the number of neurons per voxel but not the number of voxels, so it has little impact on $\alpha$ and makes memory usage grow nearly linearly with scale, without superlinear growth. This trend validates the analysis in \Cref{sec:hashmap} and demonstrates the effectiveness of the compression method.









The \smerng{} on-the-fly synapse regeneration kernel attains a peak throughput of $112$~GOPS per core. 
This method is primarily designed to compress the synaptic memory footprint, with a negligible impact on overall execution time.
Nevertheless, for workloads with intensive random number requirements, \smerng{} is theoretically capable of delivering $1.732$ EOPS computational power on the LineShine supercomputer, providing a robust foundation for stochastic computing at scale.

%% file: sections/section6_conclusion.tex
\section{Conclusion and Future Work}

The significance of CerebroSim extends beyond setting a new performance record for whole-brain simulation: it establishes a portable and broadly applicable infrastructure for extreme-scale, biologically constrained spiking brain models. This enables mechanistic studies of brain disorders through controlled in silico perturbations, offering insights into diseases such as epilepsy and neurodegeneration. In addition, the platform serves as a testbed for neuromorphic computing, facilitating the development of brain-inspired algorithms and hardware. Looking forward, it also holds promise for virtual drug screening by enabling the prediction of system-level responses to pharmacological interventions.


The broader technical implication of \myname{} is that extreme-scale brain simulation is enabled not by a single optimization but by a coordinated redesign of communication, memory access, and connectivity representation.
Delay-aware Spike Broadcast shows that delay-constrained spike propagation can be transformed into a structured and schedulable communication problem, suggesting that future communication software for sparse applications should expose aggregation, slack-aware transmission, and communication-computation overlap as first-class capabilities.
Race-free Synaptic Dynamics Computation, together with HBM placement and software prefetching, demonstrates that synchronization avoidance and locality-aware memory orchestration are essential for irregular multithreaded updates on modern compute nodes.
Sparse Synapse Storage Compression further demonstrates that trading abundant computation for on-the-fly regeneration and compressing index storage can further alleviate memory and bandwidth pressures, thereby further increasing the upper limit of sparse simulation scale.
Taken together, these three innovations provide a practical co-design template not only for brain simulation but also for other irregular and graph-like workloads, such as sparse AI, agent-based modeling, and large scientific applications with fine-grained communication and memory pressure.


These design choices translate directly into system-level performance.
On 18{,}432 LineShine nodes across 11.2 million cores, \myname{} executes a human-brain-scale model with 86 billion neurons and 100 trillion synapses, sustains 24.44~PFlop/s, and delivers 91\% weak-scaling efficiency for the core simulation phase as well as 94\% strong-scaling efficiency for the whole-brain case.
Ablation results further show that DSB reduces communication and spike-filtering overhead relative to MPI collectives, RSDC shortens the spike-delivery stage under multithreaded execution, and 3SC reduces memory usage to 65.0\% with SACI compression and to 21.8\% with RNG-based regeneration.
Measured on the full application rather than on projected kernels alone, these results indicate that \myname{} is not only fast at extreme scale but also provides a credible path toward sustained and reproducible whole-brain simulation on leadership-class systems.

However, a key challenge in whole-brain simulation lies in the current scarcity of high-resolution, large-scale neuronal recordings.
Existing experimental techniques still involve a trade-off between spatial coverage and temporal precision, and no single modality yet captures spike-level activity, structural connectivity, and whole-brain dynamics simultaneously.
Encouragingly, ongoing progress in large-scale recording, data assimilation, and hybrid modeling approaches is likely to reduce these limitations and enable more realistic system-level dynamics.
Moreover, the apparent mismatch between spiking neural network simulation outputs and fMRI signals highlights a promising direction rather than a barrier.
As neurovascular coupling models become better characterized and more tightly integrated with neural simulations, they will provide a principled bridge between fast spike-based activity and slower, spatially aggregated measurements.
In this light, the current gap between fine-grained neural simulation and coarse-grained brain observation represents an active and tractable frontier—one that is increasingly within reach as experimental and computational methods co-evolve.

%% file: sections/section93_references.tex

\bibliographystyle{IEEEtran}
\bibliography{reference/Reference}

%% file: sections/section95_biography.tex
\newpage

\section*{Biographies}

\begin{IEEEbiography}[{\includegraphics[width=1in,height=1.25in,clip,keepaspectratio]{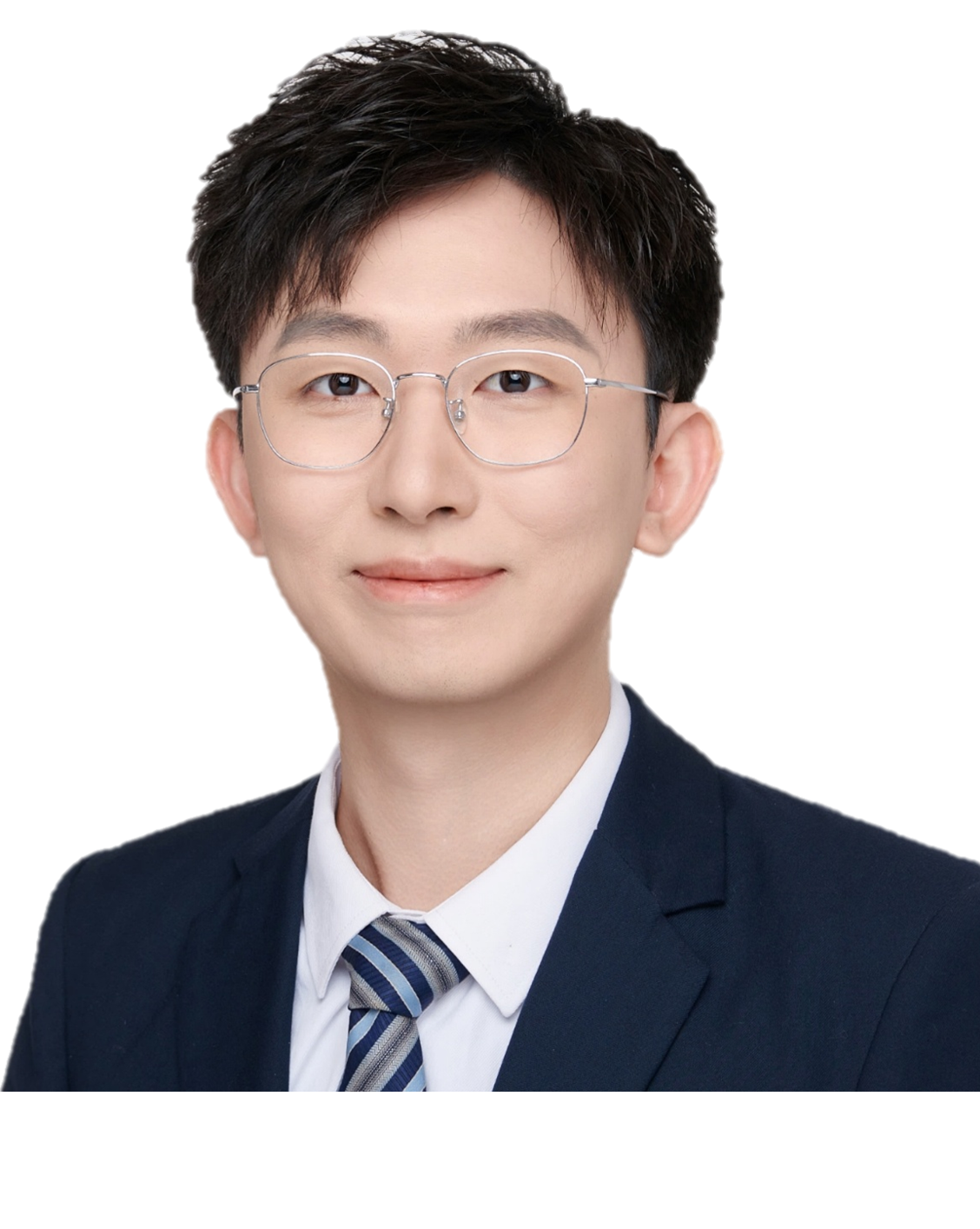}}]{Guangnan Feng}
received his B.S. and Ph.D. from Sun Yat-sen University (SYSU), Guangzhou, China, in 2019 and 2024, respectively. He is currently a postdoctoral researcher in the School of Computer Science and Engineering, SYSU. His research interests include high-performance computing, scientific computing applications, high-performance networks, and communication software.
\end{IEEEbiography}

\begin{IEEEbiography}[{\includegraphics[width=1in,height=1.25in,clip,keepaspectratio]{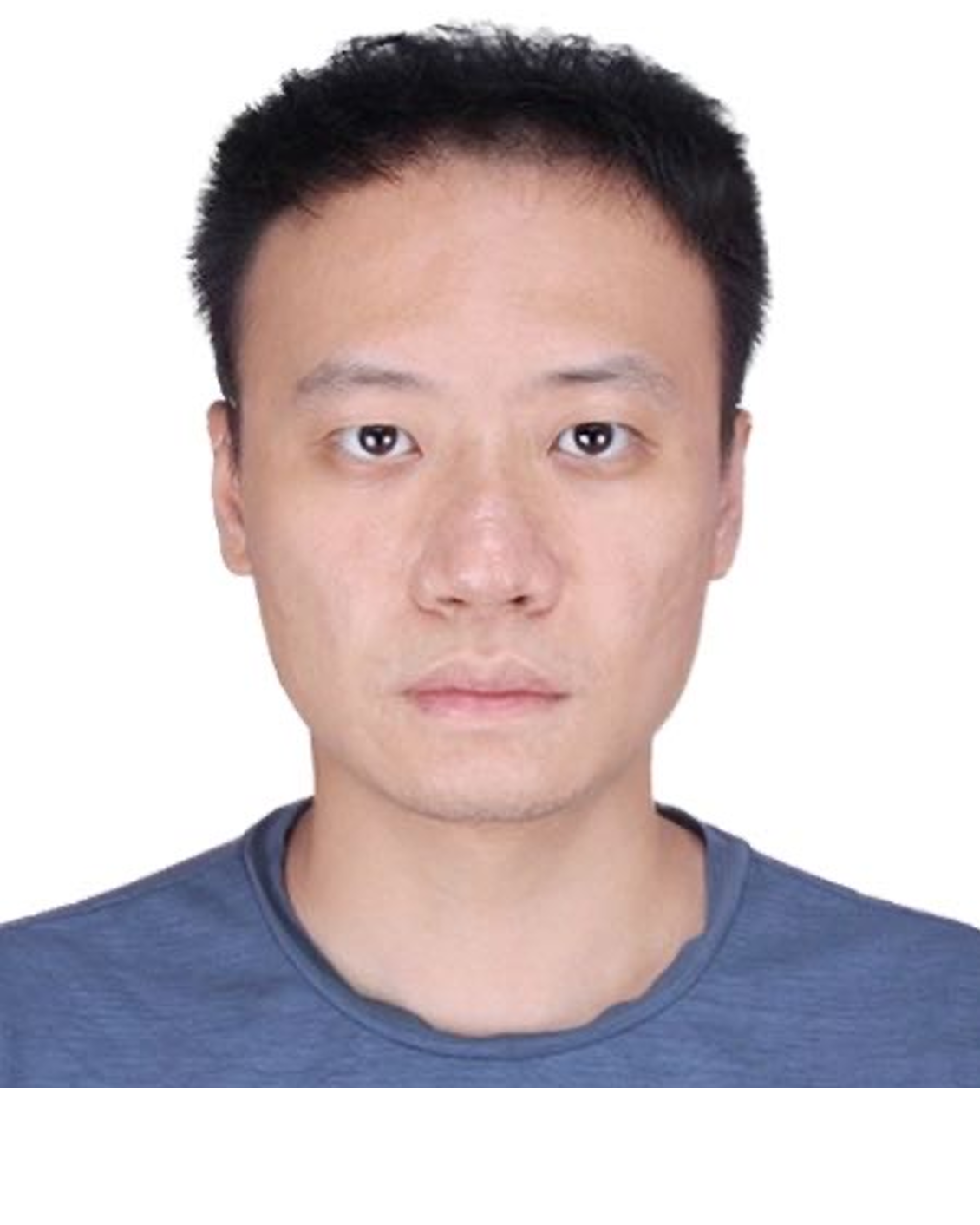}}]{Tianxiang Lyu}
received his B.S. from Sun Yat-sen University (SYSU), Shenzhen, China, in 2023 and M.S. from Juntendo University, Tokyo, Japan, in 2026, respectively. He is currently a doctoral student in the School of Computer Science and Engineering, SYSU. His research interests include high-performance computing, brain simulation, large-scale numerical analysis and quantum computing.
\end{IEEEbiography}

\begin{IEEEbiography}[{\includegraphics[width=1in,height=1.25in,clip,keepaspectratio]{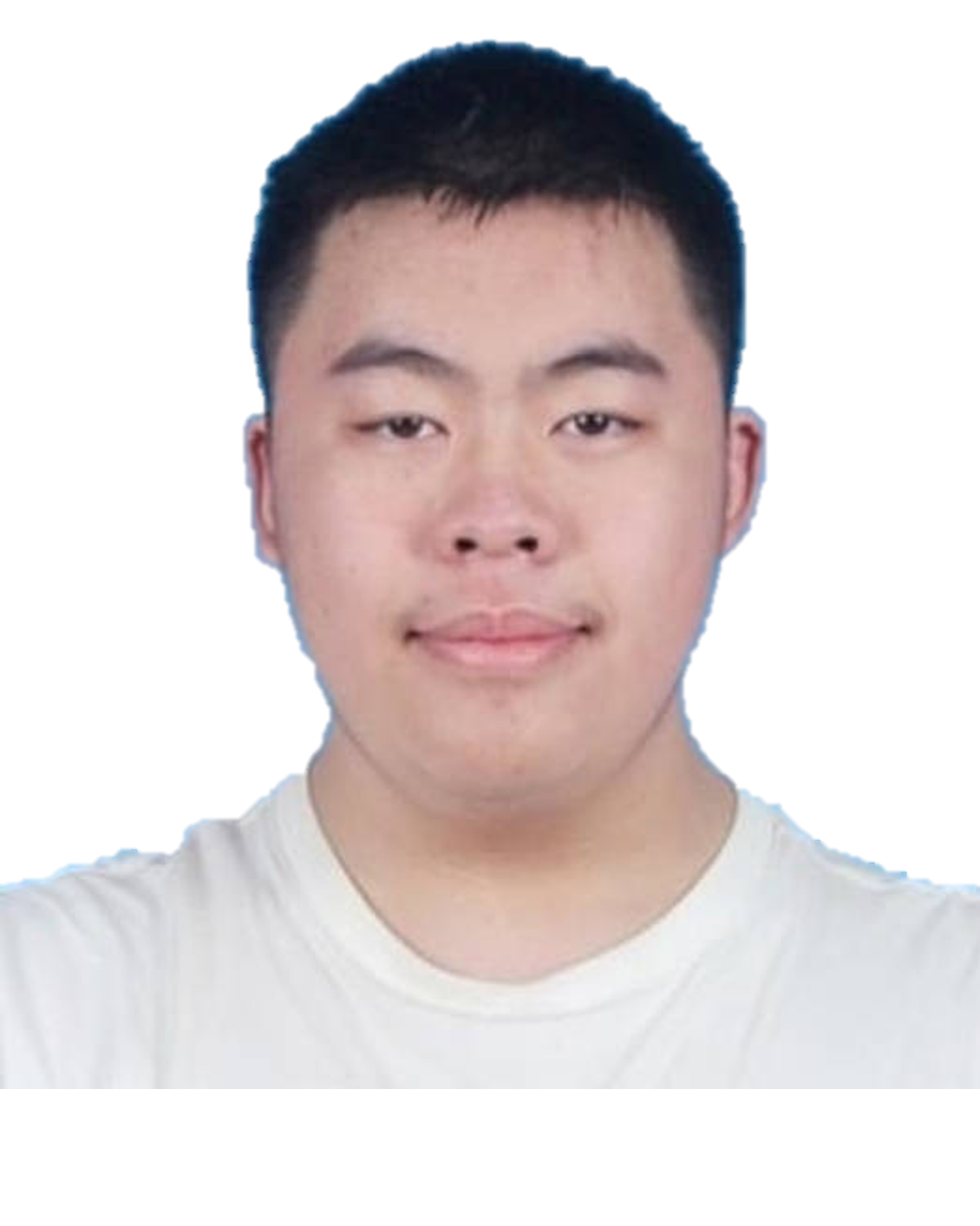}}]{Hao Huang}
received his B.S. from Chongqing University, Chongqing, China, in 2024. He is currently a graduate student in the School of Computer Science and Engineering, Sun Yat-sen University (SYSU), Guangzhou, China. His research interests include high-performance computing and large-scale brain simulation.
\end{IEEEbiography}

\begin{IEEEbiography}[{\includegraphics[width=1in,height=1.25in,clip,keepaspectratio]{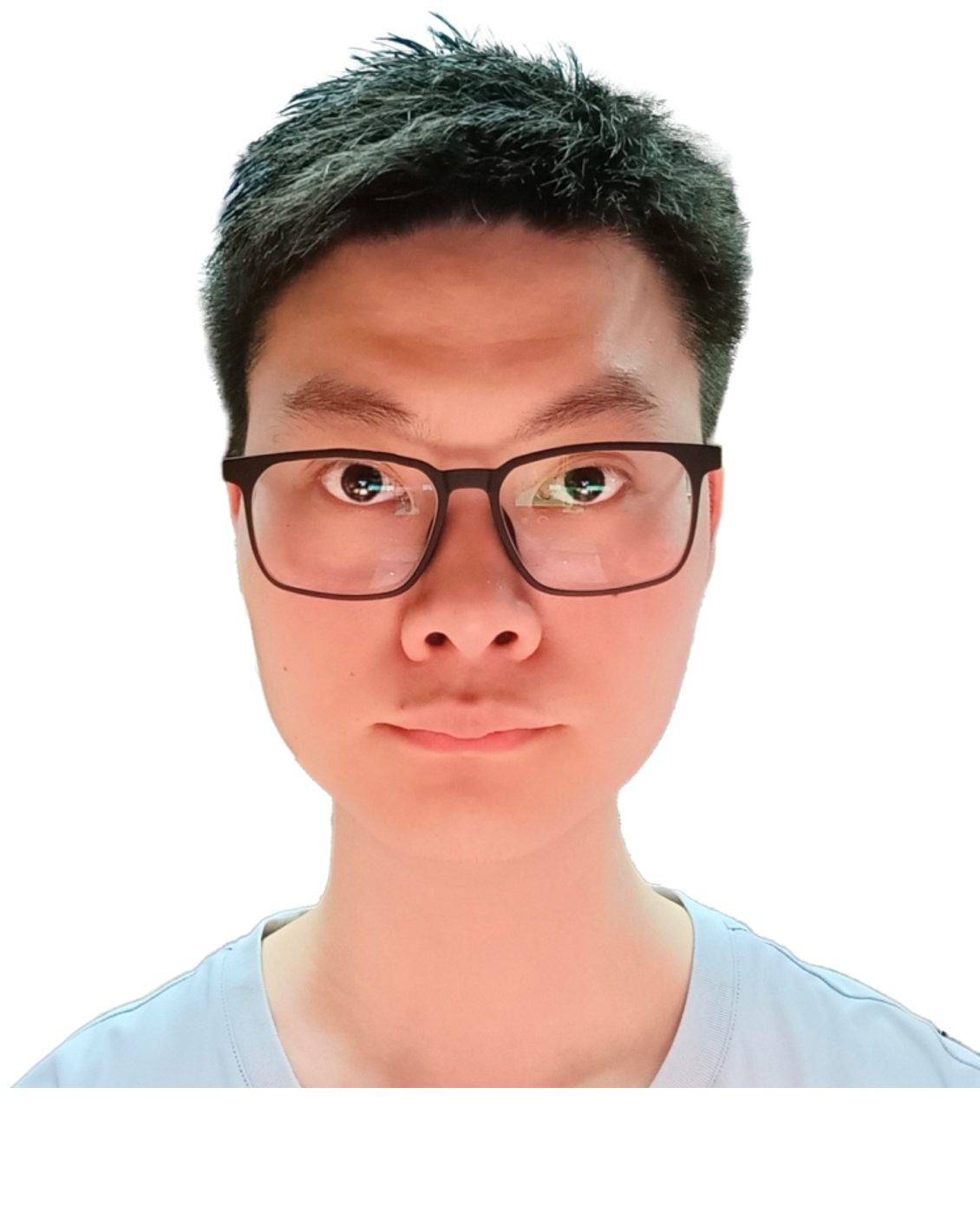}}]{Honghui Liang}
received the B.S. degree from Nankai University in 2022 and the M.S. degree from Sun Yat-sen University, Guangzhou, China, in 2025. He is currently pursuing the Ph.D. degree with the School of Computer Science, Sun Yat-sen University. His research interests include high-performance computing and brain simulation.
\end{IEEEbiography}

\begin{IEEEbiography}[{\includegraphics[width=1in,height=1.25in,clip,keepaspectratio]{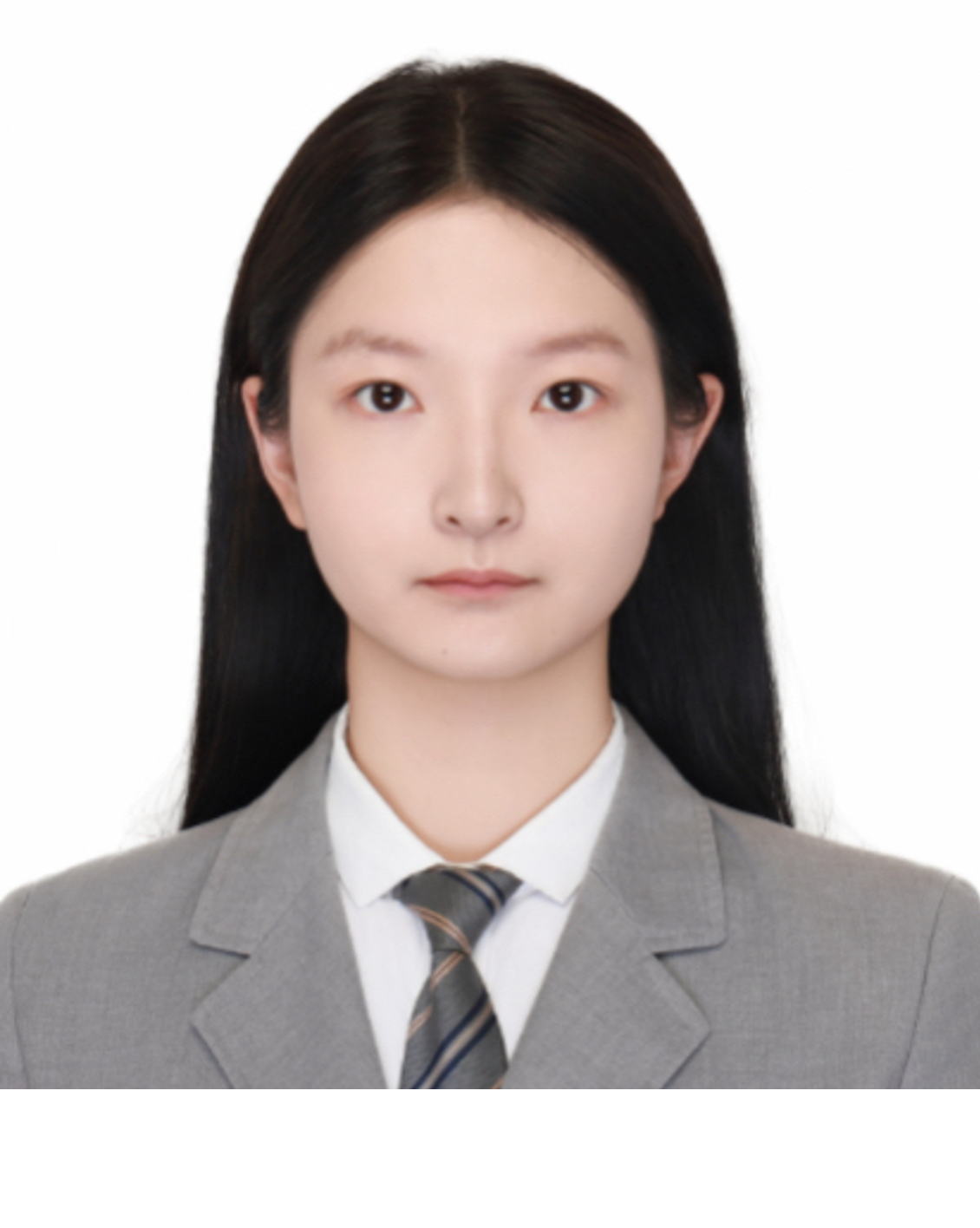}}]{Jingjing Li}
received her B.S. from Hunan University, Hunan, China, in 2025. She is currently a graduate student in the School of Computer Science and Engineering, Sun Yat-sen University (SYSU), Guangzhou, China. Her research interests include high-performance computing and large-scale brain simulation. 
\end{IEEEbiography}

\begin{IEEEbiography}[{\includegraphics[width=1in,height=1.25in,clip,keepaspectratio]{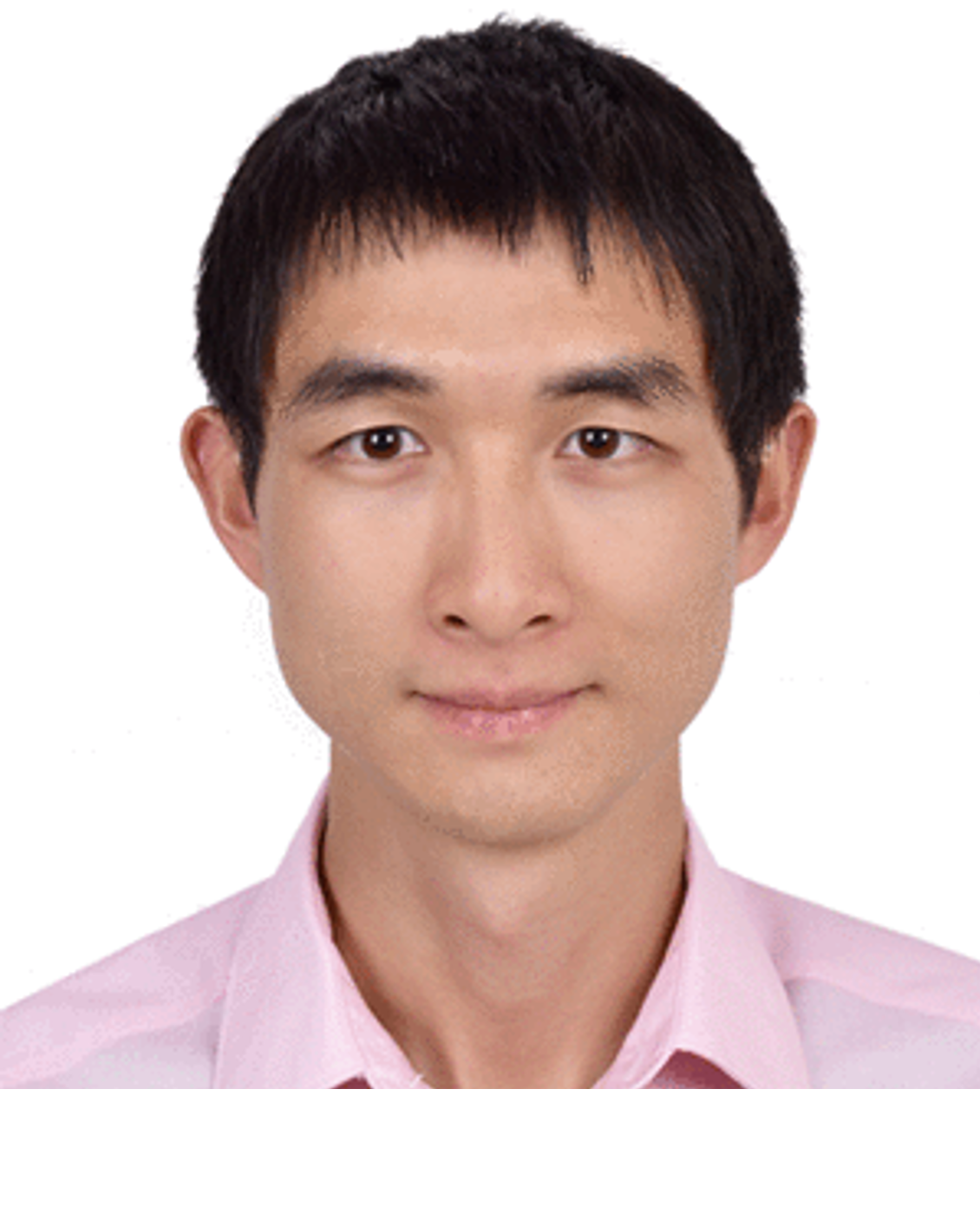}}]{Zhiguang Chen}
received the BS degree from the Harbin Institute of Technology, Harbin, and the MS and PhD degrees in computer science and technology from the National University of Defense Technology, Changsha. He is an associate professor with Sun Yat-sen University, Guangzhou. His current research interest includes distributed file system, network storage, and solid-state storage system.
\end{IEEEbiography}

\begin{IEEEbiography}[{\includegraphics[width=1in,height=1.25in,clip,keepaspectratio]{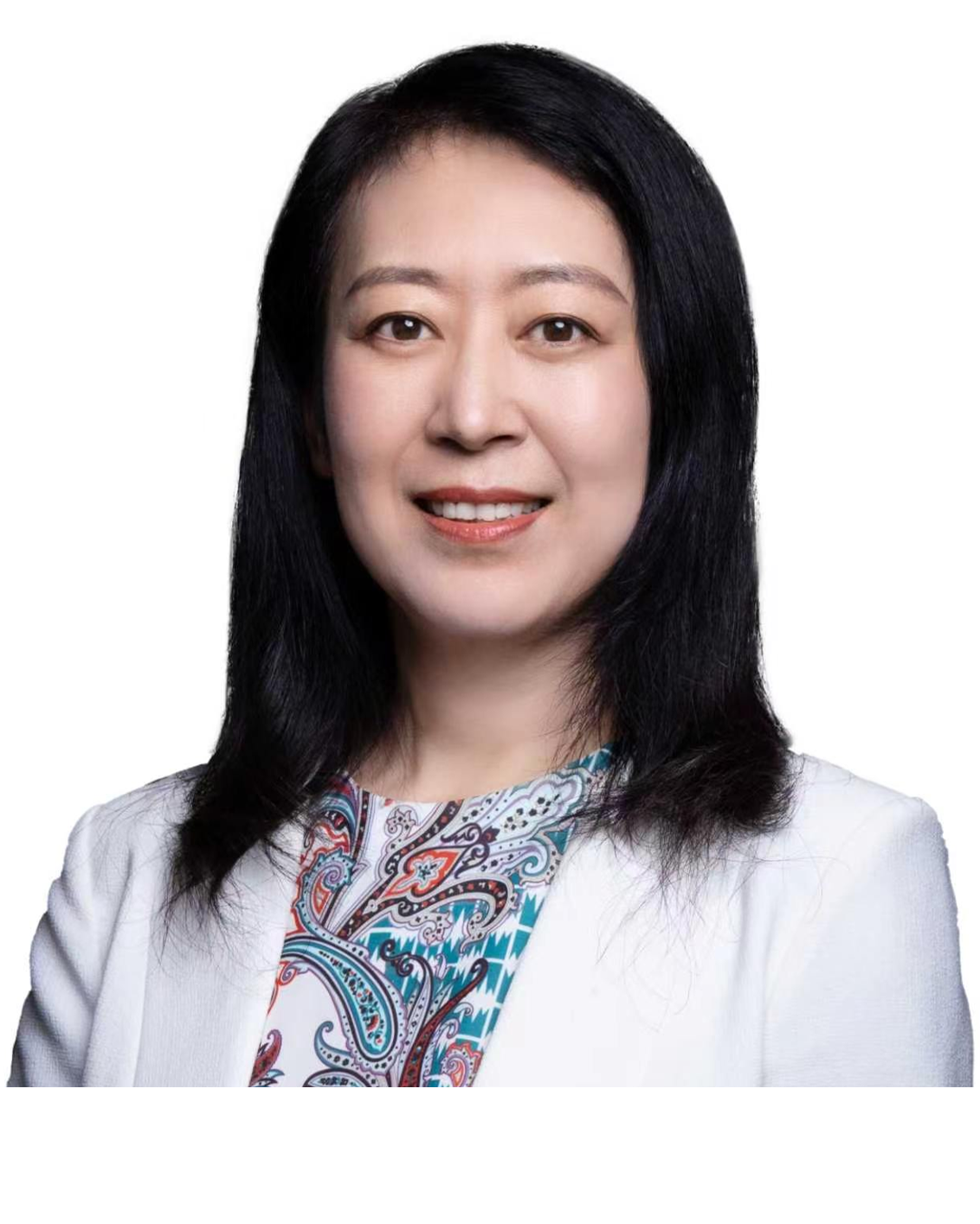}}]{Yutong Lu}
is a Professor at the School of Computer Science and Engineering, Sun Yat-sen University, China. She is the Chief Designer of the LineShine Supercomputer and serves as Director of the National Supercomputing Centers in Guangzhou and Shenzhen. She specializes in high-performance computing, with research interests spanning advanced computer architecture, programming models, and HPC-AI converged computing environments.
\end{IEEEbiography}